\documentclass{article}

\PassOptionsToPackage{numbers, compress}{natbib}

 \usepackage[final]{neurips_2019}

\usepackage[utf8]{inputenc} 
\usepackage[T1]{fontenc}    
\usepackage{hyperref}       
\usepackage{url}            
\usepackage{booktabs}       
\usepackage{amsfonts}       
\usepackage{nicefrac}       
\usepackage{microtype}      
\usepackage{algorithm}
\usepackage{algorithmic}
\usepackage{bbm}

\usepackage{amsmath}

\usepackage{color}
\usepackage{xcolor}

\usepackage[pdftex]{graphicx} 
\usepackage{graphbox}

\title{Turbo Autoencoder: Deep learning based channel codes for point-to-point communication channels}   

\author{ \parbox{1.5 in}{\centering Yihan Jiang\\
	ECE Department\\
         University of Washington\\
	Seattle, United States \\
	{\tt\small yij021@uw.edu}}
	\And
	 \parbox{1.5 in}{\centering Hyeji Kim\\
	Samsung AI Center Cambridge\\
         Cambridge, United Kingdom\\
	{\tt\small hkim1505@gmail.com}}
	\And
	 \parbox{1.5 in}{\centering Himanshu Asnani\\
	School of Technology and Computer Science\\
         Tata Institute of Fundamental Research\\
         Mumbai, India \\
	{\tt\small himanshu.asnani@tifr.res.in}}\\
	\And
	 \parbox{1.5 in}{\centering Sreeram Kannan\\
	ECE Department\\
         University of Washington\\
         Seattle, United States \\
	{\tt\small ksreeram@ee.washington.edu}}
	\And
	 \parbox{1.5 in}{\centering Sewoong Oh\\
	Allen School of Computer Science $\&$ Engineering\\
         University of Washington\\
         Seattle, United States \\
	{\tt\small sewoong@cs.washington.edu}}
	\And
	 \parbox{1.5 in}{\centering Pramod Viswanath\\
	ECE Department\\
         University of Illinois at Urbana Champaign\\
         Illinois, United States\\
	{\tt\small pramodv@illinois.edu}}
}

\begin{document}

\maketitle

\begin{abstract}

%


Designing  codes that  combat the noise in a communication medium has remained a significant area of research in information theory as well as wireless communications. Asymptotically optimal channel codes have been developed by mathematicians for communicating under canonical models after over 60 years of research. On the other hand, in many non-canonical channel settings, optimal codes do not exist and the codes designed for canonical models are adapted via heuristics to these channels and are thus not guaranteed to be optimal. In this work, we make significant progress on this problem by designing a fully end-to-end jointly trained neural encoder and decoder, namely, Turbo Autoencoder (TurboAE), with the following contributions:
($a$) under moderate block lengths, TurboAE approaches state-of-the-art performance under canonical channels; ($b$) moreover, TurboAE outperforms the state-of-the-art codes under non-canonical settings in terms of reliability. TurboAE shows that the development of channel coding design can be automated via deep learning, with near-optimal performance.



\end{abstract}

\section{Introduction}
\label{intro}
\vspace{-0.1in}
Autoencoder is a powerful unsupervised learning framework to learn latent representations by minimizing reconstruction loss of the input data~\cite{vincent2008extracting}.
Autoencoders have been widely used in unsupervised learning tasks such as representation learning~\cite{vincent2008extracting}~\cite{krizhevsky2011using}, denoising~\cite{makhzani2013k}, and generative model~\cite{chen2016infogan}~\cite{kingma2013auto}. Most autoencoders are under-complete autoencoders, for which the latent space is smaller than the input data~\cite{krizhevsky2011using}. Over-complete autoencoders have latent space larger than input data.
While the goal of under-complete autoencoder is to find a low dimensional representation of input data, the goal of over-complete autoencoder is to find a higher dimensional representation of input data so that from a noisy version of the higher dimensional representation, original data can be reliably recovered. 
Over-complete autoencoders are used in sparse representation learning~\cite{makhzani2013k}~\cite{deng2010binary} and robust representation learning~\cite{le2011ica}. 

 Channel coding aims at communicating a message over a noisy random channel~\cite{shannon1948mathematical}. As shown in Figure~\ref{ChannelAE} left, the transmitter maps a message to a codeword via adding redundancy (this mapping is called encoding). A channel between the transmitter and the receiver randomly corrupts the codeword so that the receiver observes a noisy version which is used by the receiver to estimate the transmitted message (this process is called decoding). The encoder and the decoder together can be naturally viewed as an over-complete autoencoder, where the noisy channel in the middle corrupts the hidden representation (codeword). 
 Therefore, designing a reliable autoencoder can have a strong bearing on alternative ways of designing new encoding and decoding schemes for wireless communication systems.

%
%
%
%

Traditionally, the design of communication algorithms first involves designing a `code' (i.e., the encoder) via optimizing certain mathematical properties of encoder such as minimum code distance~\cite{richardson2008modern}. The associated decoder that minimizes the bit-error-rate then is
 derived based on the maximum a posteriori (MAP) principle. However, while the optimal MAP decoder is computationally simple for some simple codes (e.g., convolutional codes), for known capacity-achieving codes, the MAP decoder is not computationally efficient; hence, alternative decoding principles such as 
  belief propagation are employed (e.g., for decoding turbo codes).  The progress on the design of optimal channel codes with computationally efficient decoders has been quite sporadic due to its reliance on human ingenuity. Since Shannon's seminal work in 1948~\cite{shannon1948mathematical}, it took several decades of research to finally reach to the current state-of-the-art codes~\cite{arikan2008channel}. 

Near-optimal channel codes such as Turbo~\cite{berrou1993near}, Low Density Parity Check (LDPC)~\cite{mackay1997near}, and Polar codes~\cite{arikan2008channel} show Shannon capacity-approaching~\cite{shannon1948mathematical} performance on AWGN channels, and they have had a tremendous impact on the Long Term Evolution (LTE) and 5G standards. 
The traditional approach has the following caveats:

($a$) Decoder design heavily relies on handcrafted optimal decoding algorithms for the canonical Additive White Gaussian Noise (AWGN) channels, where the signal is corrupted by i.i.d. Gaussian noise. 
In practical channels, when the channel deviates from AWGN settings, often times heuristics are used to compensate the non-Gaussian properties of the noise, which leaves a room for the potential improvement in reliability of a decoder~\cite{richardson2008modern}~\cite{li2013ofdma}.

($b$) 
Channel codes are designed for a finite block length $K$. 
Channel codes are guaranteed to be optimal only when the block-length approaches infinity, and  thus are near-optimal in practice only when the block-length is large. On the other hand, under short and moderate block length regimes, there is a room for improvement~\cite{polyanskiy2010channel}.

($c$) The encoder designed for the AWGN channel is used across a large family of channels, while the decoder is adapted. This design methodology fails to utilize the flexibility of the encoder.


\noindent
{\bf Related work.}
Deep learning has pushed the state-of-the-art performance of computer vision and natural language processing to a new level far beyond handcrafted algorithms in a data-driven fashion~\cite{goodfellow2016deep}. 
There also has been a recent movement in applying deep learning to wireless communications. 
Deep learning based channel decoder design has been studied since~\cite{o2016learning}~\cite{o2017introduction}, where encoder is fixed as a near-optimal code. It is shown that belief propagation decoders for LDPC and Polar codes can be imitated by neural networks~\cite{nachmani2016learning}~\cite{nachmani2018deep}~\cite{xu2017improved}~\cite{gruber2017deep}~\cite{cammerer2017scaling}. It is also shown that convolutional and turbo codes can be decoded optimally via Recurrent Neural Networks (RNN)~\cite{kim2018communication} and Convolutional Neural Networks (CNN)~\cite{jiang2019deepturbo}. Equipping a decoder with a learnable neural network also allows fast adaptation via meta-learning~\cite{jiang2019mind}. Recent works also extend deep learning to multiple-input and multiple-output (MIMO) settings~\cite{samuel2017deep}.  While {\em neural decoders} show improved performance on various communication channels, 
 there has been limited success in inventing novel codes using this paradigm. Training methods for improving both modulation and channel coding are introduced in~\cite{o2016learning}~\cite{o2017introduction}, where a (7,4) neural code mapping a 4-bit message to a length-7 codeword can match (7,4) Hamming code performance. 
Current research includes training an encoder and a decoder with noisy feedback~\cite{aoudia2018end}, improving modulation gain~\cite{felix2018ofdm}, as well as extensions to multi-terminal settings~\cite{OShea-Erpek-Clancy2017}. Joint source-channel coding shows improved results combining source coding (compression) along with channel coding (noise mitigation)~\cite{choi2018necst}.  Neural codes were shown to outperform existing state-of-the-art codes on the feedback channel~\cite{kim2018deepcode}. However, in the canonical  setting of AWGN channel, neural codes are still far from capacity-approaching performance due to the following challenges.

(Challenge A) Encoding with randomness is critical to harvest coding gain on long block lengths~\cite{shannon1948mathematical}. However, existing sequential neural models, both CNN and even RNN, can only learn limited local dependency~\cite{chung2014empirical}. Hence, neural encoder cannot sufficiently utilize the benefits of even moderate block length. 

(Challenge B) Training neural encoder and decoder jointly (with a random channel in between) introduces optimization issues where the algorithm gets stuck at local optima. Hence, a novel training algorithm is needed.


\noindent
{\bf Contributions.} 
In this paper, we confront the above challenges by introducing Turbo Autoencoder (henceforth, TurboAE) -- the first channel coding scheme with both encoder and decoder powered by neural networks that achieves reliability close to the state-of-the-art channel codes under AWGN channels for a moderate block length. 
We demonstrate that channel coding, which has been a focus of study by mathematicians for several decades~\cite{richardson2008modern}, can be learned in an end-to-end fashion from data alone. 
Our major contributions are:

\begin{itemize} 
	\item  We introduce TurboAE, a neural network based over-complete autoencoder parameterized as Convolutional Neural Networks (CNN) along with interleavers (permutation) and de-interleavers (de-permutation) inspired by turbo codes (Section~\ref{turboAE}). 
	We introduce TurboAE-binary, which binarizes the codewords via straight-through estimator (Section~\ref{binaryTurboAE}). 
	\item We propose techniques that are critical for training TurboAE which includes mechanisms of alternate training of encoder and decoder as well as strategies to choose right training examples. 
	Our training methodology  ensures stable training of TurboAE without getting trapped at locally optimal encoder-decoder solutions.
(Section~\ref{sec:training})
\item Compared to multiple capacity-approaching codes on AWGN channels, TurboAE shows superior performance in the low to middle SNR range when the block length is of moderate size ($K\sim100$). To the best of our knowledge, this is the first result demonstrating the deep learning powered discovered neural codes can outperform traditional codes in the canonical AWGN setting (Section~\ref{sec:expawgn}).
	\item On a non-AWGN channel, fine-tuned TurboAE shows significant improvements over state-of-the-art coding schemes due to the flexibility of encoder design, which shows that TurboAE has advantages on designing codes where handcrafted solutions fail (Section~\ref{sec:expnonawgn}).	
\end{itemize}

We make our source codes public available in https://github.com/yihanjiang/turboae/, and refer the interested readers to appendix for more detailed design and performances.


\section{Problem Formation}
\label{problem_form}
\begin{figure}[!h] 
\centering
\includegraphics[width=0.12\textwidth]{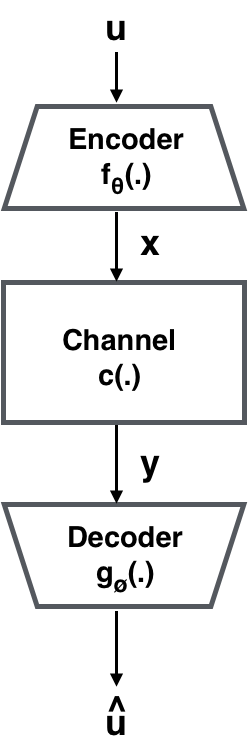}\ \ \ 
\includegraphics[width=0.80\textwidth]{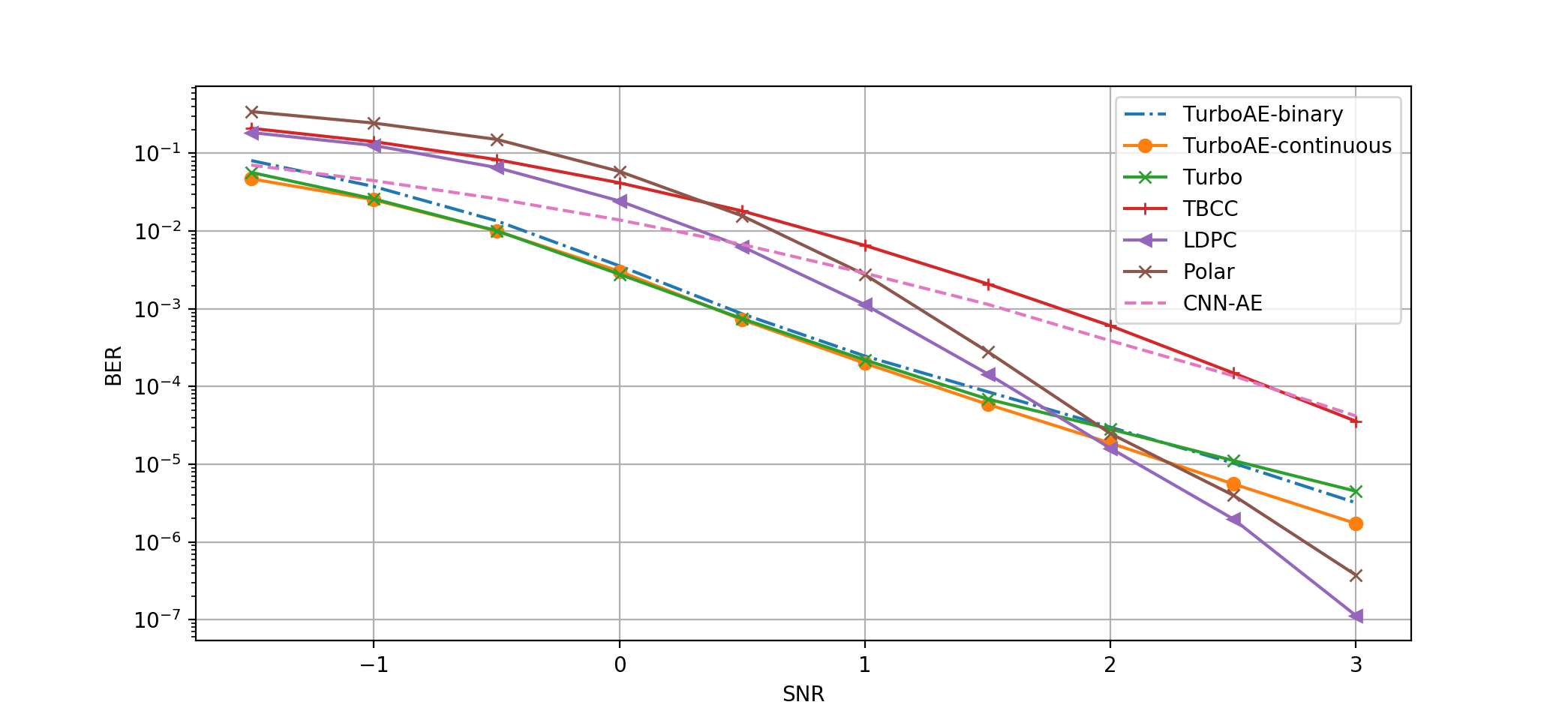}\ \ \ 
\caption{Channel coding can be viewed as an over-complete autoencoder with channel in the middle (left). TurboAE performs well under moderate block length in low and middle SNR (right).
}\label{ChannelAE} 
\centering
\end{figure}

\vspace{-0.1in}
The channel coding problem is illustrated in Figure~\ref{ChannelAE} left, which consists of three blocks -- an encoder $f_\theta(\cdot)$, a channel $c(\cdot)$, and a decoder $g_\phi(.)$. A channel $c(\cdot)$ randomly corrupts an input $x$ and is represented as a probability transition function $p_{y|x}$. A canonical example of channel $c(\cdot)$ is an identically and independently distributed (i.i.d.) AWGN channel, which generates $y_i = x_i + z_i$ for $z_i \sim N(0, \sigma^2)$, $i = 1, \cdots, K$. The encoder $x = f_\theta(u)$ maps a random binary message sequence $u = (u_1, \cdots, u_K) \in \{0,1\}^K$ of block length $K$ to a codeword $x=(x_1, \cdots, x_N)$ of length N, where $x$ must satisfy either soft power constraint where $E(x)=0$ and $E(x^2)=1$, or hard power constraint $x \in \{-1, +1\}$. Code rate is defined as $R = \frac{K}{N}$, where $N > K$. The decoder $g_\phi(y)$ maps a real valued received sequence $y = (y_1, \cdots, y_N) \in \mathcal{R}^N$ to an estimate of the transmitted message sequence $\hat{u} = (\hat{u}_1, \cdots, \hat{u}_K) \in \{0,1\}^K$. 

AWGN channel allows closed-form mathematical analysis, which has remained as the major playground for channel coding researchers. The noise level is defined as signal-to-noise ratio, $SNR = -10\log_{10} \sigma^2$. The decoder recovers the original message as $\hat u = g_\phi(y)$ using the received signal $y$. 

Channel coding aims to minimize the error rate of recovered message $\hat u$. The standard metrics are bit error rate (BER), defined as $BER=\frac{1}{K}\sum_1^K\mathrm{Pr}(\hat u_i \neq u_i)$, and block error rate (BLER), defined as $BLER = \mathrm{Pr}(\hat u \neq u)$.

While canonical capacity-approaching channel codes work well as block length goes to infinity, when the block length is short, they are not guaranteed to be optimal. We show the benchmarks on block length 100 in Figure~\ref{ChannelAE} right with widely-used LDPC, Turbo, Polar, and Tail-bitting Convolutional Code (TBCC), generated via Vienna 5G simulator~\cite{Vienna5GSLS}~\cite{tahir2017ber}, with code rate $1/3$. 

Naively applying deep learning models by replacing encoder and decoder with general purpose neural network does not perform well.  Direct applications of fully connected neural network (FCNN) cannot scale to a longer block length; the performance of FCNN-AE is even worse than repetition code~\cite{jiang2018learn}. 
Direct applications where both the encoder and the decoder are Convolutional Autoencoder (termed as CNN-AE~\cite{zhu2018joint}) shows better performance than TBCC, but are far from capacity-approaching codes such as LDPC, Polar, and Turbo.  Bidirectional RNN and LSTM ~\cite{jiang2018learn} has similar performance as CNN-AE and is not shown in the figure for clarity. Thus neither CNN nor RNN  based auto-encoders can directly approach state-of-the-art performance. A key reason for their shortcoming is that they have only local memory, the encoder only remembers information locally. To have high protection against channel noise, it is necessary to have long term memory. 


We propose TurboAE with interleaved encoding and iterative decoding that creates long term memory in the code and shows a significant improvement compared to CNN-AE. TurboAE has two versions, TurboAE-continuous which faces soft power constraint (i.e., the total power across a codeword is bounded) and TurboAE-binary which faces hard power constraint (i.e., each transmitted symbol has a power constraint - and is thus forced to be binary). Both TurboAE-binary and TurboAE-continuous perform comparable or better than all other capacity-approaching codes at a low SNR, while at a high SNR (over 2 dB with $BER<10^{-5}$), the performance is only worse than LDPC and Polar code. 

\section{TurboAE : Architecture Design and Training} 
\label{turbofy}
\subsection{Design of TurboAE}\label{turboAE}

{\bf Turbo code and turbo principle}: 
Turbo code is the first capacity-approaching code ever designed~\cite{berrou1993near}. There are two novel components of Turbo code which led to its success: an interleaved encoder and an iterative decoder. The starting point of the Turbo code is a recursive systematic convolutional (RSC) code which has an optimal decoding algorithm (the Bahl-Cocke-Jelinek-Raviv (BCJR) algorithm~\cite{bahl1974optimal}). A key disadvantage in the RSC code is that the algorithm lacks long range memory (since the convolutional code operates on a sliding window). The key insight of Berrou was to introduce long range memory by creating two copies of the input bits - the first goes through the RSC code and the second copy goes through an interleaver (which is a permutation of the bits) before going through the same code. Such a code can be decoded by iteratively alternating between soft-decoding based on the signal received from the first copy and then using the de-interleaved version as a prior to decode the second copy. The `Turbo principle' ~\cite{hagenauer1997turbo} refers to the iterative decoding with successively refining the posterior distribution on the transmitted bits across decoding stages with original and interleaved order. This code is known to have excellent performance, and inspired by this, we design TurboAE featuring both learnable interleaved encoder and iterative decoder.

\begin{figure}[!h] 
\centering
\includegraphics[width=0.5\textwidth]{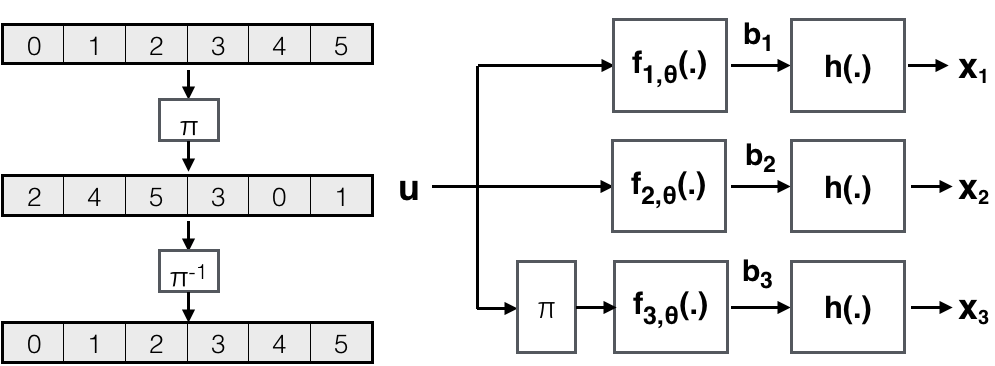}\ \ \ 
\caption{Visualization of Interleaver ($\pi$) and De-interleaver ($\pi^{-1}$) (left); TurboAE encoder on code rate 1/3 (right)}\label{enc_inter}
\centering
\end{figure}

{\bf Interleaved Encoding Structure}:
Interleaving is widely used in communication systems to mitigate bursty noise ~\cite{sadjadpour2001interleaver}. Formally, interleaver $x^{\pi} = \pi(x)$ and de-interleaver $x = \pi^{-1}(x^{\pi} )$ shuffle and shuffle back the input sequence $x$ with the a pseudo random interleaving array known to both encoder and decoder, respectively, as shown in Figure~\ref{enc_inter} left. In the context of Turbo code and TurboAE, the interleaving is not used to mitigate bursty errors (since we are mainly concerned with i.i.d. channels) but rather to add long range memory in the structure of the code. 

We take code rate 1/3 as an example for interleaved encoder $f_{\theta}$, which consists of three learnable encoding blocks $f_{i, \theta}(.)$ for $i \in \{1,2,3\}$, where $f_{i, \theta}(.)$ encodes $b_i = f_{\theta}(u), i \in \{1,2\}$ and $b_3 = f_{3, \theta}(\pi(u))$, where $b_i$ is a continuous value. The power constraint of channel coding is enforced via power constraint block $x_i = h(b_i)$.

\begin{figure}[!h] 
\centering
\includegraphics[width=0.95\textwidth]{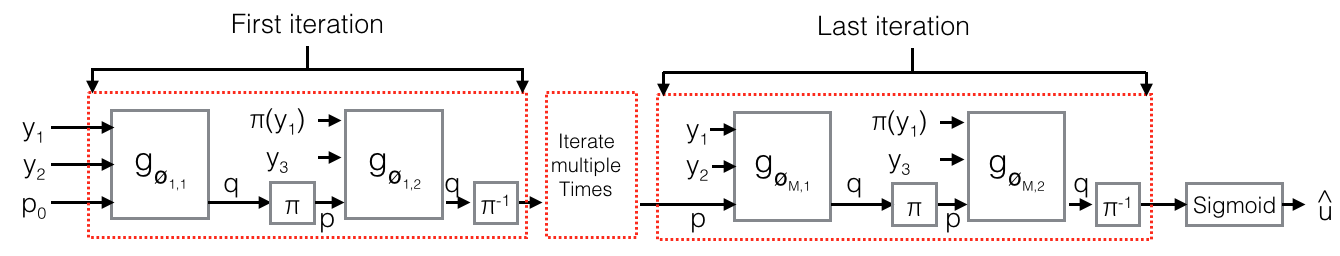}\ \ \ 
\caption{TurboAE iterative decoder on code rate 1/3}\label{turboprinciple}
\centering
\end{figure}

{\bf Iterative Decoding Structure}: 
As received codewords are encoded from original message $u$ and interleaved message $\pi(u)$, decoding interleaved code requires iterative decoding on both interleaved and de-interleaved order shown in Figure \ref{turboprinciple}. Let $y_1, y_2,y_3$ denote noisy versions of $x_1, x_2, x_3$, respectively. The decoder runs multiple iterations, with each iteration contains two decoders $g_{\phi_{i,1}}$ and $g_{\phi_{i,2}}$ for interleaved and de-interleaved order on the $i$-th iteration.

The first decoder $g_{\phi_{i,1}}$ takes received signal $y_1$, $y_2$ and de-interleaved prior $p$ with shape $(K,F)$, where as $F$ is the information feature size for each code bit, to produce the posterior $q$ with same shape $(K,F)$.
The second decoder $g_{\phi_{i,2}}$ takes interleaved signal $\pi(y_1)$, $y_3$ and interleaved prior $p$ to produce posterior $q$.
The posterior of previous stage $q$ serves as the prior of next stage $p$. 
The first iteration takes 0 as a prior, and at last iteration the posterior is of shape $(K, 1)$, are decoded as by sigmoid function as $\hat u = sigmoid(q)$. 

Both encoder and decoder structure can be considered as a parametrization of Turbo code. Once we parametrize the encoder and the decoder, since the encoder, channel, and decoder are differentiable, TurboAE can be trained end-to-end via gradient descent and its variants.

{\bf Encoder and Decoder Design}:
The space of messages and codewords are exponential (For a length-$K$ binary sequence, there are $2^K$ distinct messages). Hence, the encoder and decoder must have some structural restrictions to ensure generalization to messages unseen during the training~\cite{dorner2017deep}. Applying parameter-sharing sequential neural models such as CNN and RNN are natural parametrization methods for both the encoding and the decoding blocks.

RNN models such as Gated Recurrent Unit (GRU) and Long-Short Term Memory (LSTM) are commonly used for sequential modeling problems~\cite{bahdanau2014neural}. RNN is widely used in deep learning based communications systems~\cite{kim2018communication}~\cite{kim2018deepcode}~\cite{jiang2019deepturbo}~\cite{jiang2018learn}, as RNN has a natural connection to sequential encoding and decoding algorithms such as convolutional code and BCJR algorithm~\cite{kim2018communication}. 

However RNN models are: (1) of higher complexity than CNN models, (2) harder to train due to gradient explosion, and (3) harder to run in parallel~\cite{chung2014empirical}. In this paper, we use one dimensional CNN (1D-CNN) as the alternative encoding and decoding model. Although the longest dependency length is fixed, 1D-CNN has lower complexity, better trainability~\cite{yin2017comparative}, and easier implementation in parallel via AI-chips~\cite{ovtcharov2015accelerating}. The learning curve comparison between CNN and RNN is shown in Figure~\ref{cnn_vs_rnn} left. Training CNN-based model converges faster and more stable than RNN-based GRU model. 
The TurboAE complexity is shown in appendix.

{\bf Power Constraint Block}\label{ste}:
The operation of power constraint blocks (i.e., $h(\cdot)$ in $x = h(b)$) depends on the requirement of power constraint.

Soft power constraint normalize the power of code, as $E(x)=0$ and $E(x^2)=1$. TurboAE-continuous with soft power constraint allows the code $x$ to be continuous. Addressing the statistical estimation issue given a limited batch size, we use normalization method~\cite{lei2016layer} as:$x_i = \frac{b_i -\mu(b)}{\sigma(b)}$,
where $\mu(b) = \frac1K \sum_{i=1}^{K}{b_i}$ and $\sigma(b)=\sqrt{\frac1K \sum_{i=1}^{K}(b_i -\mu(b))^2}$ are scalar mean and standard deviation estimation of the whole block. During the training phase, $\mu(b)$ and $\sigma(b)$ are estimated from the whole batch. On the other hand, during the testing phase, $\mu(b)$ and $\sigma(b)$ are pre-computed with multiple batches. The normalization layer can be also considered as BatchNorm without affine projection, which is critical to stabilize the training of the encoder~\cite{santurkar2018does}.

\subsection{Design of TurboAE-binary -- Binarization via Straight-Through Estimator}\label{binaryTurboAE}
Some wireless communication system requires a hard power constraint, where the encoder output is binary as $x \in \{-1,+1\}$~\cite{tse2005fundamentals} - so that every symbol has exactly the same power and the information is conveyed in the sign. Hard power constraint is not differentiable, since restricting $x \in \{-1,+1\}$ via $x = \mathrm{sign}(b)$ has zero gradient almost everywhere. We combine normalization and Straight-Through Estimator (STE)~\cite{bengio2013estimating}~\cite{hubara2016binarized} to bypass this differentiability issue. 
STE passes the gradient of $x = \mathrm{sign}(b)$ as $\frac{\partial x}{\partial b} =\mathbbm{1} (|b| \leq 1)$ and enables training of an encoder by passing estimated gradients to the encoder, while enforcing hard power constraint. 

\begin{figure}[!h] 
\centering
\includegraphics[width=0.95\textwidth]{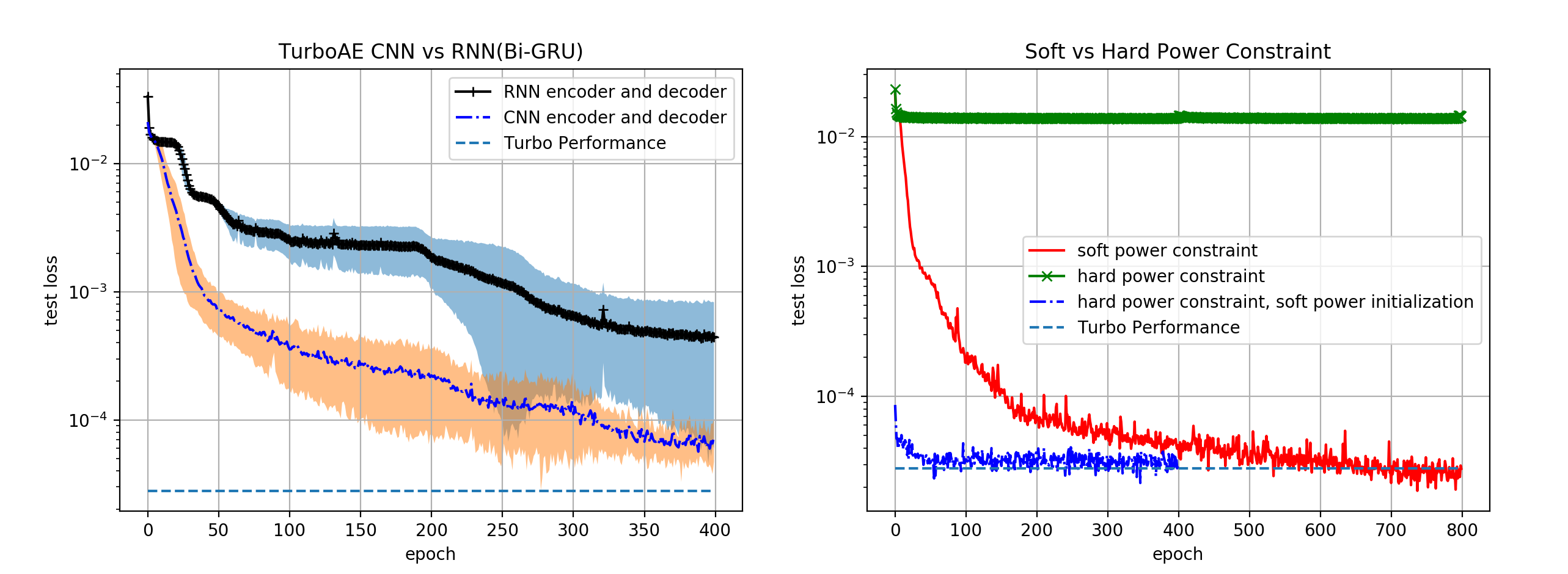}\ \ \ 
\caption{Learning Curves on CNN vs GRU: CNN shows faster training convergence (left); Training with STE requires soft-constraint pre-training (right)}\label{cnn_vs_rnn}
\centering
\end{figure}

Simply training with STE cannot learn a good encoder as shown in Figure~\ref{cnn_vs_rnn} right. To mitigate the trainability issue, we apply pre-training, which pre-trains TurboAE-continuous firstly, and then add the hard power constraint on top of soft power constraint as $x = \mathrm{sign}(\frac{b -\mu(b) }{\sigma(b)})$, whereas the gradient is estimated via STE. Figure~\ref{cnn_vs_rnn} right shows that with pre-training, TurboAE-binary reaches Turbo performance within 100 epochs of fine-tuning.

TurboAE-binary is slightly worse than TurboAE-continuous as shown in Figure~\ref{ChannelAE}, especially at high SNR, since: ($a$) TurboAE-continuous can be considered as a joint coding and high order modulation scheme, which has a larger capacity than binary coding at high SNR~\cite{tse2005fundamentals}, and ($b$) STE is an estimated gradient, which makes training encoder more noisy and less stable.

\subsection{Neural Trainability Design}\label{sec:training}
The training algorithms for training TurboAE are shown in Algorithm 1. Compared to the conventional deep learning model training, training TurboAE has the following differences:


\begin{algorithm}
\caption{Training Algorithm for TurboAE}
\begin{algorithmic} 
\REQUIRE Batch Size $B$, Train Encoder Steps $T_{enc}$, Train Decoder Steps $T_{dec}$, Number of Epoch $M$
Encoder Training SNR $\sigma_{enc}$, Decoder Training SNR $\sigma_{dec}$

\FOR{$i \leq M$} 
\FOR{$j \leq T_{enc}$} 
\STATE Generate random training example $u$, and random noise $z \sim N(0, \sigma_{enc})$.
\STATE Train encoder $f_{\theta}$ with decoder fixed, with $u$ and $z$.
\ENDFOR

\FOR{$j \leq T_{dec}$} 
\STATE Generate random training example $u$, and random noise $z \sim N(0, \sigma_{dec})$.
\STATE Train decoder $g_{\phi}$ with encoder fixed, with $u$ and $z$.
\ENDFOR

\ENDFOR
\end{algorithmic}\label{alg}
\end{algorithm}

\begin{itemize}
\item \textbf{Very Large Batch Size} Large batch size is critical to average the channel noise effects. Empirically, TurboAE reaches Turbo performance only when the batch size is grater than 500.

\item \textbf{Train Encoder and Decoder Separately} We train encoder and decoder separately as shown in Algorithm 1, to avoid getting stuck in local optimum~\cite{aoudia2018end}~\cite{jiang2018learn}.


\item \textbf{Different Training Noise Level for Encoder and Decoder} 
Empirically, while it is best to train a decoder at a low training SNR as discussed in~\cite{kim2018communication}, it is best to train an encoder at a training SNR that matches the testing SNR, e.g training encoder at 2dB results in good encoder when testing at 2dB~\cite{jiang2018learn}. In this work, we use random selection of -1.5 to 2 dB for training the decoder, and test and train the encoder at the same SNR. 

\end{itemize}

We do a detailed analysis of training algorithms in the supplementary materials. The hyper-parameters are shown in Table 1.
\begin{table}[!h] 
\centering
{\small
\begin{tabular}{ |p{4.5cm}|p{9cm}|  }
 \hline
 Loss  & Binary Cross-Entropy (BCE)\\
 Encoder   & 2 layers 1D-CNN, kernel size 5, 100 filters for each $f_{i,\theta}(.)$ block\\
 Decoder  & 5 layers 1D-CNN, kernel size 5, 100 filters for each $g_{\phi_{i,j}(.)}$ block\\
 Decoder Iterations   & 6 \\
 Info Feature Size F   & 5 \\
 Batch Size & 500 when start, double when saturates for 20 epochs, till reaches 2000\\
 Optimizer   & Adam with initial learning rate 0.0001\\
 Training Schedule for Each Epoch & Train encoder $T_{enc}=100$ times, then train decoder $T_{dec}=500$ times\\ 
 Block Length K  & 100 \\
 Number of Epochs M & 800\\
 \hline
\end{tabular}\\
\label{table:1}
\vspace{0.1in}
}\caption{Hyper-parameters of TurboAE}
\end{table}

\section{Experiment Results}\label{sec:exp}

\subsection{Block length coding gain of TurboAE}\label{sec:expawgn} 
\begin{figure}[!h] 
\centering
\includegraphics[width=0.80\textwidth]{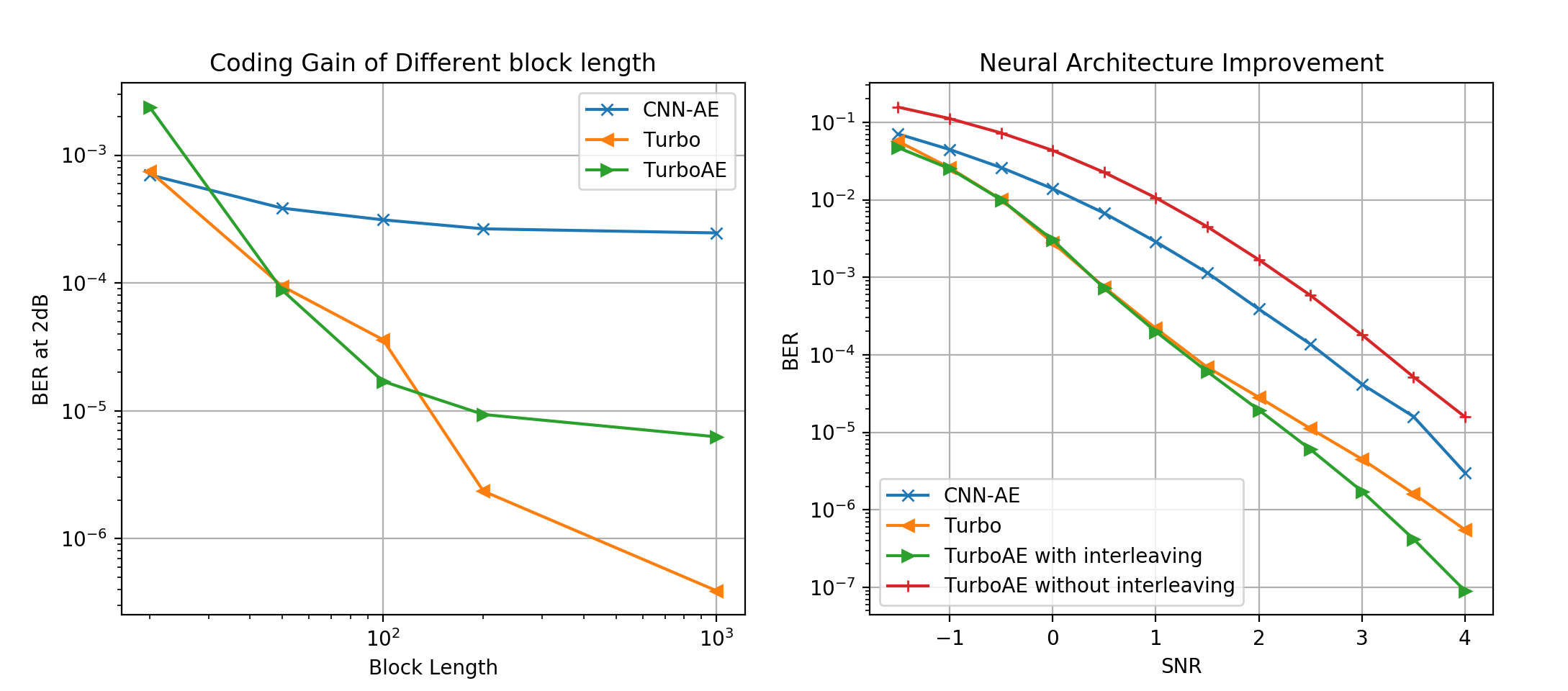}\ \ \ 
\caption{Interleaving improves blocklength gain (left); Neural Architecture improves BER performance (right).}\label{code_gain}
\centering 
\end{figure}

As block length increases, better reliability can be achieved via channel coding, which is referred to as \emph{blocklength gain} ~\cite{berrou1993near}.  We compare TurboAE (only TurboAE-continuous is shown in this section) with the Turbo code and CNN-AE, tested at BER at 2dB on different block lengths, shown in Figure \ref{code_gain} left. Both CNN-AE and TurboAE are trained with block length 100, and tested on various block lengths. As the block length increases, CNN-AE shows saturating blocklength gain, while TurboAE and Turbo code reduce the error rate as the block length increases. Naively applying general purpose neural network such as CNN to channel coding problem cannot gain performance on long block lengths.

Note that TurboAE  is still worse than Turbo when the block length is large, since long block length requires large memory usage and more complicated structure to train. Improving TurboAE on very long block length remains open as an interesting future direction.

The BER performance boosted by neural architecture design is shown in Figure~\ref{code_gain} right. We compare the fine-tuned performance among CNN-AE, TurboAE, and TurboAE without interleaving as $x^{\pi} = \pi(x)$. TurboAE with interleaving significantly outperforms TurboAE without interleaving and CNN-AE. 

\subsection{Performance on non-AWGN channels}\label{sec:expnonawgn}

\begin{figure}[!h] 
\centering
\includegraphics[width=0.80\textwidth]{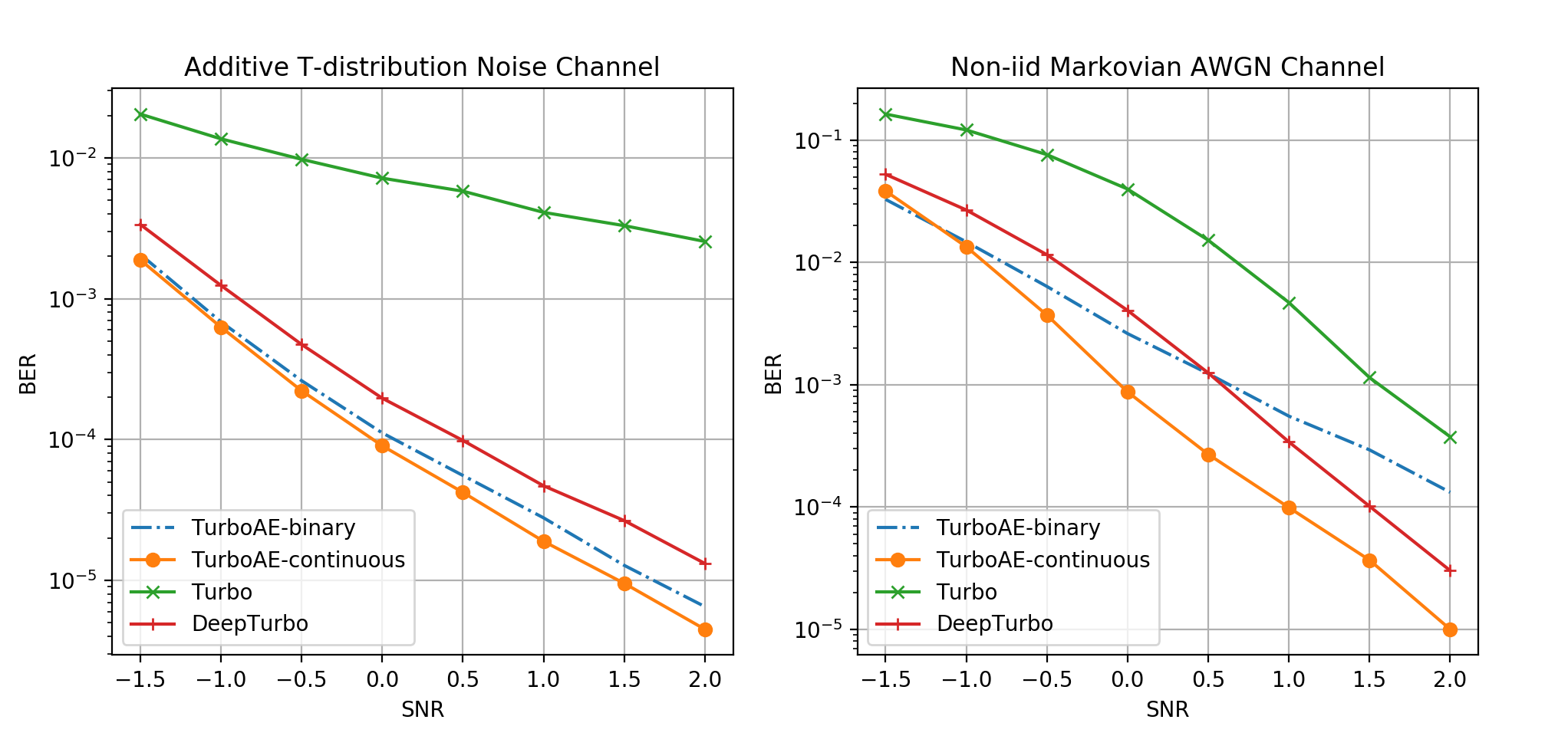}\ \ \ 

\caption{TurboAE on iid ATN channel (left) and on-iid Markovian-AWGN channel (right)}\label{nonawgn}
\centering
\end{figure}

Typically there are no close-form solutions under non-AWGN and non-iid channels. We compare two benchmarks: (1) canonical Turbo code, and (2) DeepTurbo Decoder~\cite{jiang2019deepturbo}, a neural decoder fine-tuned at the given channel. We test the performance on both iid channels and non-iid channels in settings as follows:

($a$) iid Additive T-distribution Noise (ATN) Channel, with $y_i = x_i + z_i$, where iid $z_i \sim T(\nu, \sigma^2)$ is heavy-tail (tail weight controlled based on the parameter $\nu=3.0$) T-distribution noise with variance $\sigma^2$. The performance is shown in Figure~\ref{nonawgn} left.

($b$) non-iid Markovian-AWGN channel, is a special AWGN channel with two states, $\{$good, bad$\}$. 
 At bad state the noise is worse by 1dB than the SNR, and at good state, the noise is better by 1dB than the SNR. The state transition probability between good and bad states are symmetric as $p_{bg} = p_{gb}= 0.8$. The performance is shown in Figure~\ref{nonawgn} right.

For both ATN and Markovian-AWGN channels, DeepTurbo outperforms canonical Turbo code. TurboAE-continuous with learnable encoder outperforms DeepTurbo in both cases. TurboAE-binary outperforms DeepTurbo on ATN channel, while on Markovian-AWGN channel, TurboAE-binary does not perform better than DeepTurbo at high SNR regimes (but still outperforms canonical Turbo). With the flexibility of designing an encoder, TurboAE designs better code than handcrafted Turbo code, for channels without a closed-form mathematical solution. 

\section{Conclusion and discussion}\label{conclusion}
In summary, in this paper, we propose TurboAE, an end-to-end learnt channel coding scheme with novel neural structure and training algorithms. TurboAE learns capacity-approaching code on various channels under moderate block length by building upon `turbo principle' and thus, exhibits discovery of codes for channels where a closed-form representation may not exist. TurboAE, hence, brings an interesting research direction to design channel coding algorithms via joint encoder and decoder design.

A few pending issues hamper further improving TurboAE. \textbf{Large block length} requires extensive training memory. 
With enough computing resources, we believe that TurboAE's performance at larger block lengths can potentially improve. 
\textbf{High SNR} training remains hard, as in high SNR the error events become extremely rare. \textbf{Optimizing BLER} requires novel and stable objective for training. Such pending issues are interesting future directions.

\subsubsection*{Acknowledgments}
This work was supported in part by NSF awards 1908003, 651236 and
1703403.

\medskip

\small

\bibliographystyle{IEEEtran}
\bibliography{nips2019}

\newpage
\appendix
\section{TurboAE Design Analysis}
\subsection{Neural Architecture Design}
\subsubsection{Supporting code rates beyond 1/3}
\begin{figure}[!h] 
\centering
\includegraphics[width=0.95\textwidth]{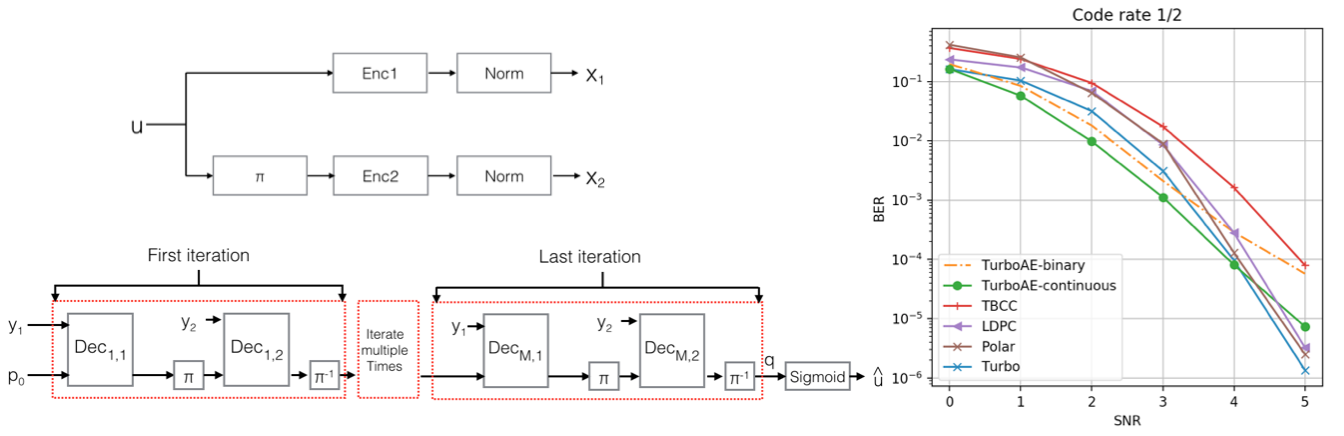}\ \ \ 
\caption{The encoder structure (up left), decoder structure (down left), and BER performance (right) of code rate 1/2}\label{rate2}
\centering
\end{figure}

In main text, only neural code for code rate $R=1/3$ is shown. The TurboAE encoder and decoder for code rate 1/2 is shown in Figure~\ref{rate2}. Still designed under `Turbo principle', TurboAE with code rate 1/2 shows impressive performance under low to moderate SNR, within block length 100. To generate code rates beyond 1/2, we can utilize puncturing.

\subsubsection{CNN with Residual Connection}

\begin{figure}[!h] 
\centering
\includegraphics[width=0.8\textwidth]{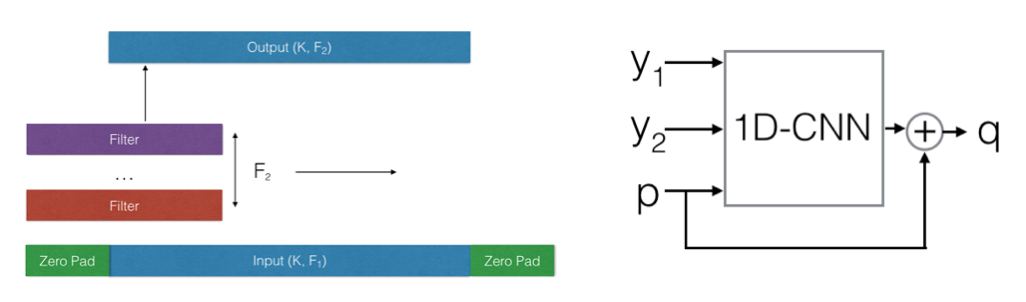}\ \ \ 

\caption{1D CNN visualization on 1 layer (left); CNN with residual connection (right).}\label{cnn_fig}
\centering
\end{figure}

The same shape property of 1D-CNN is preserved by setting odd kernel size $k$ equals twice the zero-padding length minus one, as shown in Figure~\ref{cnn_fig} left.
The encoder simply use 1D-CNN as encoder blocks, while the decoder uses residual connection to bypass gradient on iterative decoding procedure to improve trainability~\cite{he2016deep}, and also inspired by extrinsic information from Turbo code~\cite{hagenauer1997turbo}, shown in Figure \ref{cnn_fig} right. Adding residual connection improves training speed and improve final BER performance~\cite{jiang2019deepturbo}.

\subsubsection{Network Size}
\begin{figure}[!h] 
\centering
\includegraphics[width=0.45\textwidth]{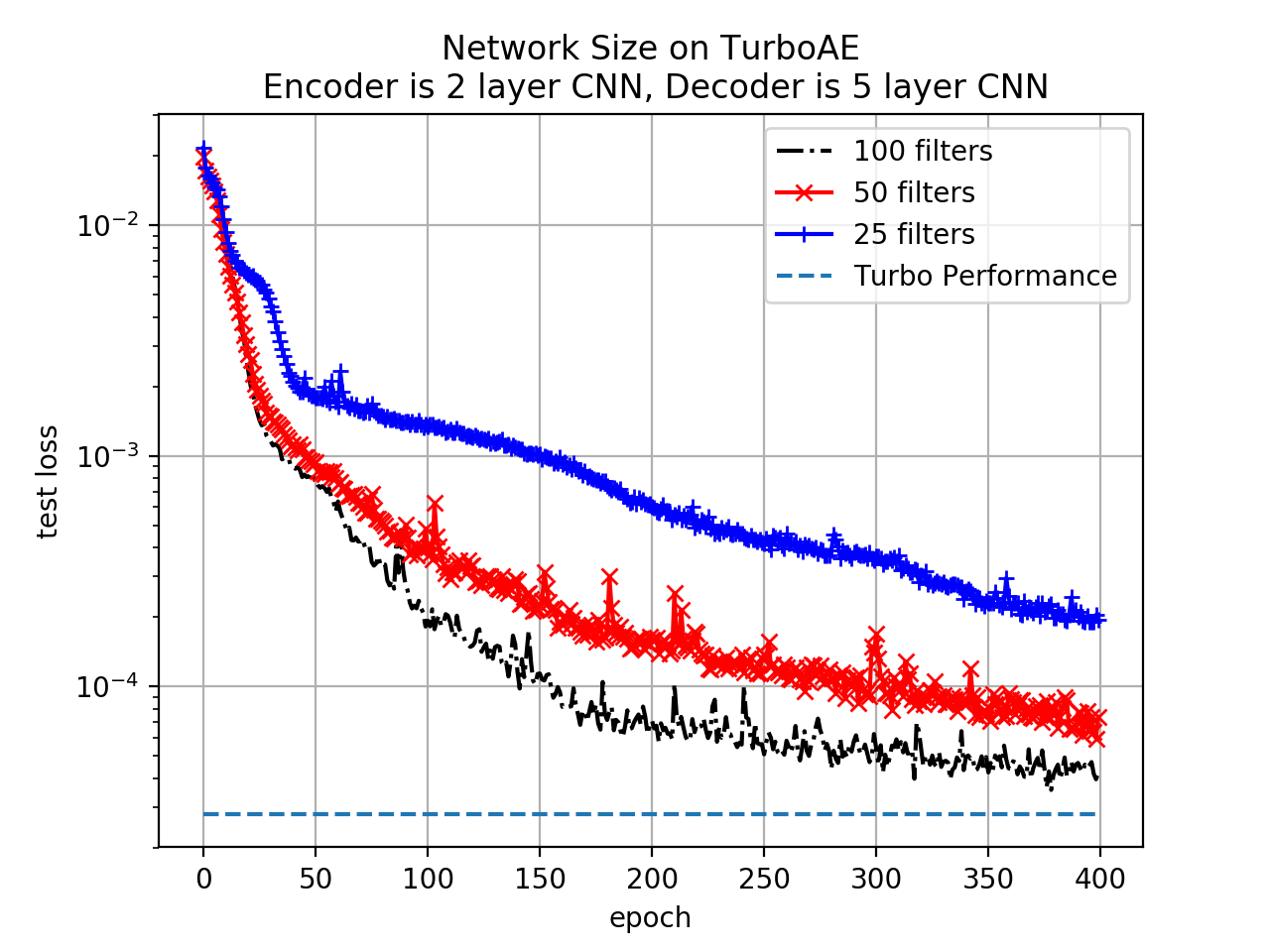}\ \ \ 
\includegraphics[width=0.45\textwidth]{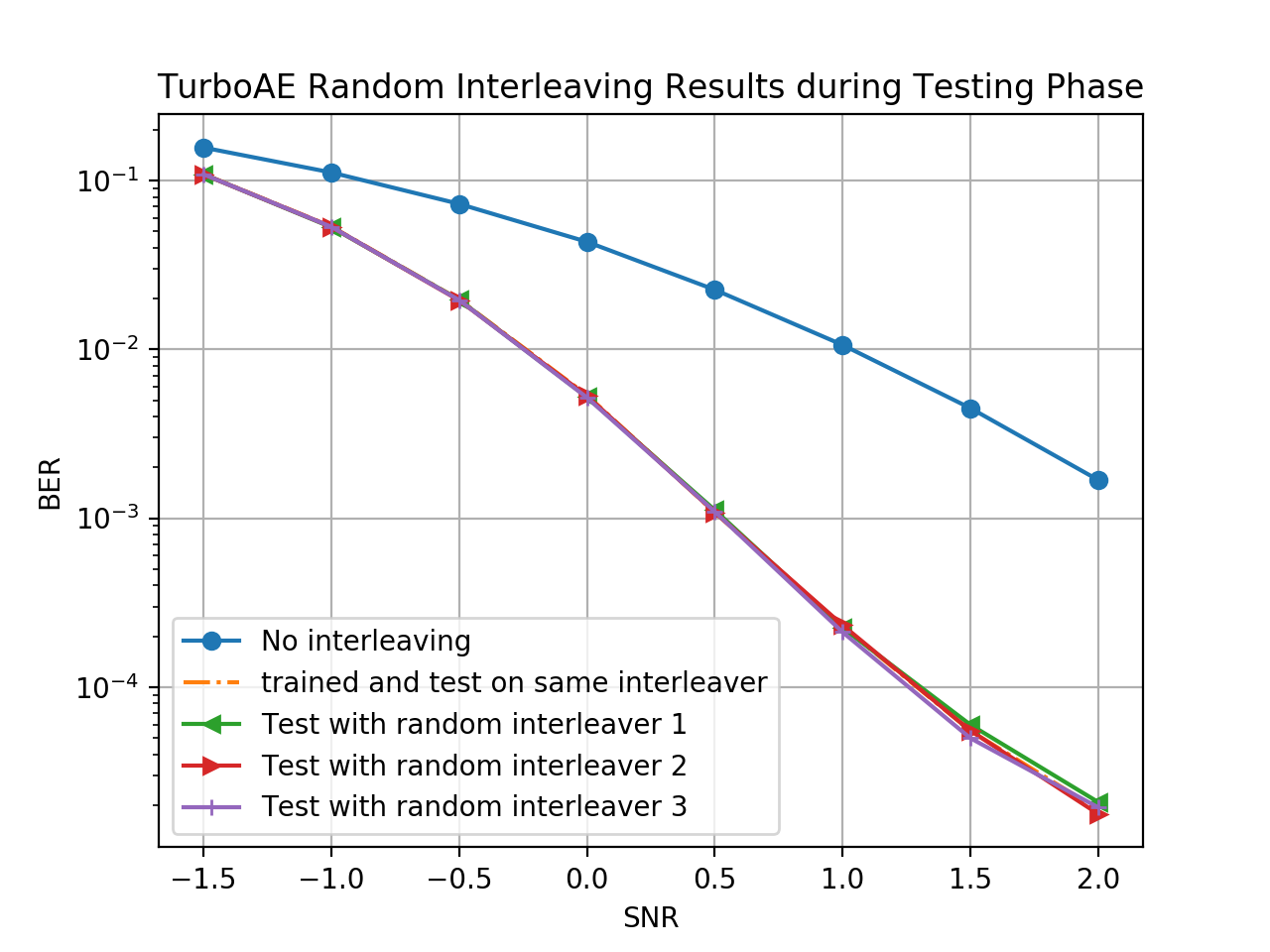}\ \ \ 
\caption{Larger Network has better performance (left); Random interleaving array shows same performance (right).}\label{nw_size}
\centering
\end{figure}

In figure \ref{nw_size} left, we show the test loss trajectory of TurboAE with different network size. We keep both encoder and decoder with same number of filters. Larger network lead to faster training and better performance, with the cost of larger computation and memory usage. We take encoder and decoder with 100 filters, which trains fast given limited computational resource (e.g., training 400 epochs takes 1 day on one Nvidia 1080Ti.)

\subsubsection{Random Interleaving Array During Testing Phase}
Given a fixed pseudo-random interleaving array, one concern is that TurboAE could overfit to specific interleaving array, and when both encoder and decoder change the interleaving array, TurboAE will have a degraded performance. However, empirically, we observe that TurboAE doesn't overfit to the training fixed pseudo-random interleaving array, as shown in Figure \ref{nw_size} right. The TurboAE is trained on one specific interleaving array, and tested on 3 random generated interleaving arrays. For TurboAE, whenever the interleaving array is pseudo-random, the neural encoder and decoder still learn without overfitting.

However, when the interleaving array is not random, e.g not applying interleaving as $y = \pi(x)$, termed as `no interleaving', the performance degrades significantly. 

\subsection{Training Algorithms}

\subsubsection{Joint Training vs Separate Training}
\begin{figure}[!h] 
\centering
\includegraphics[width=0.9\textwidth]{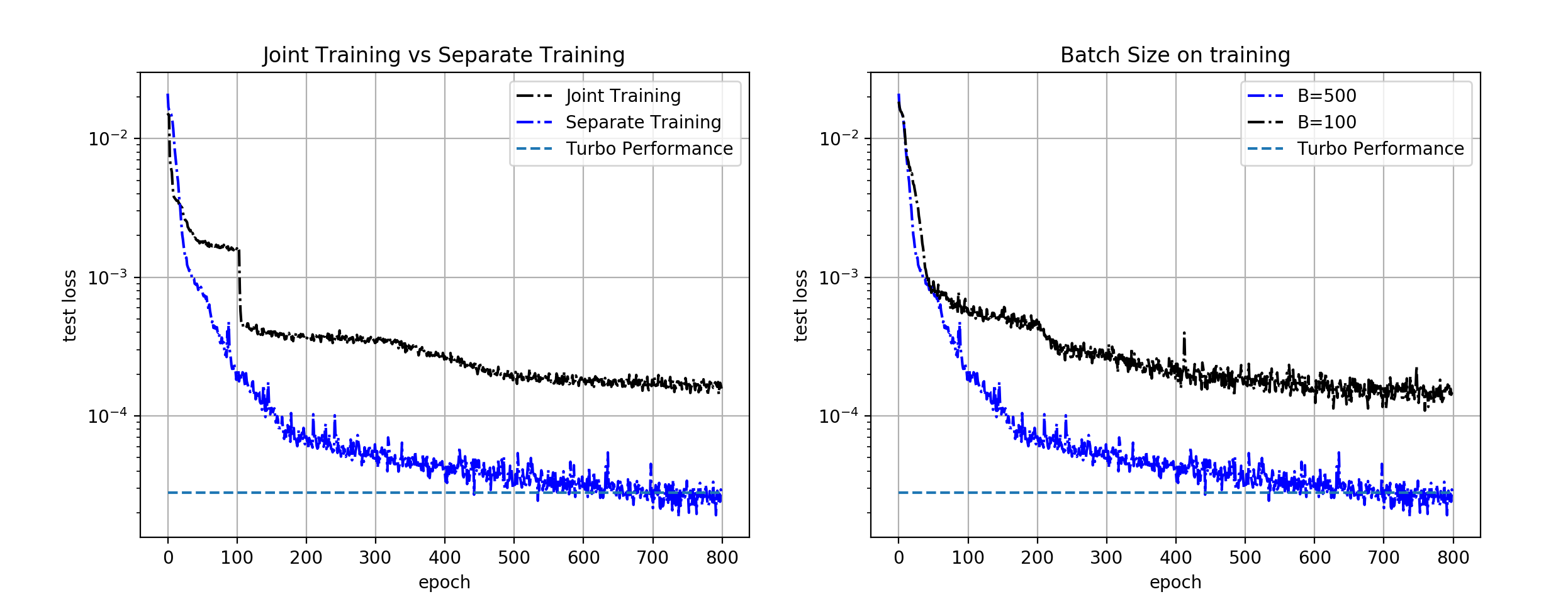}\ \ \ 
\caption{Training encoder and decoder jointly gets stuck as local optimum (left). Large batch size improves training (right).}\label{joint}
\centering
\end{figure}

Empirically training encoder and decoder simultaneously is easier to get stuck in local optimum as shown in Figure \ref{joint} left. Training encoder and decoder separately is less likely to get stuck in local optimum~\cite{aoudia2018end}~\cite{jiang2018learn}. Training decoder more times than encoder, on the other hand, makes decoder better approximates optimal decoding algorithm of the encoder, which offers more accurate estimated gradient and stabilizes the training process~\cite{jiang2018learn}. We training encoder and decoder separately, with each epoch trains encoder 100 times and decoder 500 times. 

\subsubsection{Large Batch Size Improves Training Significantly}\label{bs}
Large batch size helps training deep generative models such as Generative Adversarial Networks (GAN)~\cite{goodfellow2014generative} and Variantional Autoencoder (VAE)~\cite{kingma2013auto}, and is also critical to training TurboAE. Figure~\ref{joint} right shows that large batch size leads to significantly lower test BER.

The analysis is on AWGN channel by using the 1st order Taylor expansion on decoder $g_{\phi}(.)$ as: $\hat u = g_{\phi}(x+z) = g_{\phi}(x)+ z g'_{\phi}(x) + O(z^2)$.

Taking gradient of both sides becomes:
$\frac{\partial \hat u }{\partial x} \approx g'_{\phi}(x)+ z g''_{\phi}(x)$ 
and
$\frac{\partial \hat u}{\partial \phi} \approx \frac{\partial g_{\phi}(x)}{\partial \phi}+ z \frac{\partial g'_{\phi}(x)}{\partial \phi}$ 

AWGN channel has $\frac{\partial y}{\partial x} = 1$ with iid noise. Consider the normalization layer $x = h(b)$, the gradient pass through normalization layer with batch size $B$ is~\cite{santurkar2018does}:

\begin{equation}
\frac{\partial x_i}{\partial b_j} = \frac{1}{\sigma(b)} (\mathbbm{1} (i=j) - \frac1B (1+b_i b_j))
\label{enc_norm_grad}
\end{equation}

Known $\hat u = sigmoid(q)$, as $q = g_\phi(h(f_\theta(u))+z)$, the gradient of BCE loss with respect to logit $q$ is $\frac{\partial BCE(u, \hat u)}{\partial q}  = \hat u - u $, the gradient of encoder is:
\begin{equation}
\frac{\partial BCE(u, \hat u)}{\partial \theta} = \frac{\partial BCE(u, \hat u)}{\partial q}  \frac{\partial q}{\partial y} \frac{\partial y}{\partial x} \frac{\partial x}{\partial b} \frac{\partial b}{\partial \theta} =  (\hat u-u) (g'_{\phi}(x)+ z g''_{\phi}(x))  \frac{\partial x}{\partial b}   \frac{\partial f_{\theta}(u)}{\partial \theta}
\label{enc_grad}
\end{equation}

The gradient of decoder is:
\begin{equation}
\frac{\partial L}{\partial \phi} = \frac{\partial BCE(u, \hat u)}{\partial q} \frac{\partial q}{\partial \phi}  = (\hat u-u) (\frac{\partial g_{\phi}(x)}{\partial \phi}+ z \frac{\partial g'_{\phi}(x)}{\partial \phi})
\label{dec_grad}
\end{equation}

The benefits of large batch size are as follows:
\begin{itemize}
\item \textbf{Less noisy gradient for encoder.} The gradient passes through normalization layer is as shown in Equation (\ref{enc_norm_grad}). 
With large batch size $B$, the gradient passes through normalization reduces the noise introduced by $\frac1B (1+b_i b_j)$, making the gradient passed to encoder less noisy.

\item \textbf{Larger batch size reduces gradient noise.} Large batch size makes gradient estimation for both encoder and decoder more accurate, as the error term $z g''_{\phi}(x)$  in Equation (\ref{enc_grad}) and $z \frac{\partial g'_{\phi}(x)}{\partial \phi}$ in Equation (\ref{dec_grad}) can be reduced with large batch size with expectation $E[z] = 0$. Better gradients for both encoder and decoder improve training. 

\item \textbf{More accurate statistics for normalization.} With larger batch size, the mean and the standard deviation for normalization used in power normalization are more accurate, which introduces less noise.

\end{itemize}

\subsubsection{Training SNR} 
Training noise level (SNR) is an critical parameter for training TurboAE. The training SNR analysis can be derived by the gradient analysis of section \ref{bs}. The training noise has different effect on encoder and decoder. The training noise affects decoder with noise term $z \frac{\partial g'_{\phi}(x)}{\partial \phi}$in Equation (\ref{dec_grad}). Given an fixed encoder, training decoder with different SNR results in different levels of regularization. For encoder there are two source of noise regularizations: ($a$) $z g''_{\phi}(x)$  in Equation (\ref{enc_grad},) and ($b$) noise introduced by normalization layer in Equation (\ref{enc_norm_grad}). Training encoder with different SNR also results in different levels of regularization, which differs from training decoders. 

\begin{figure}[!h] 
\centering
\includegraphics[width=0.45\textwidth]{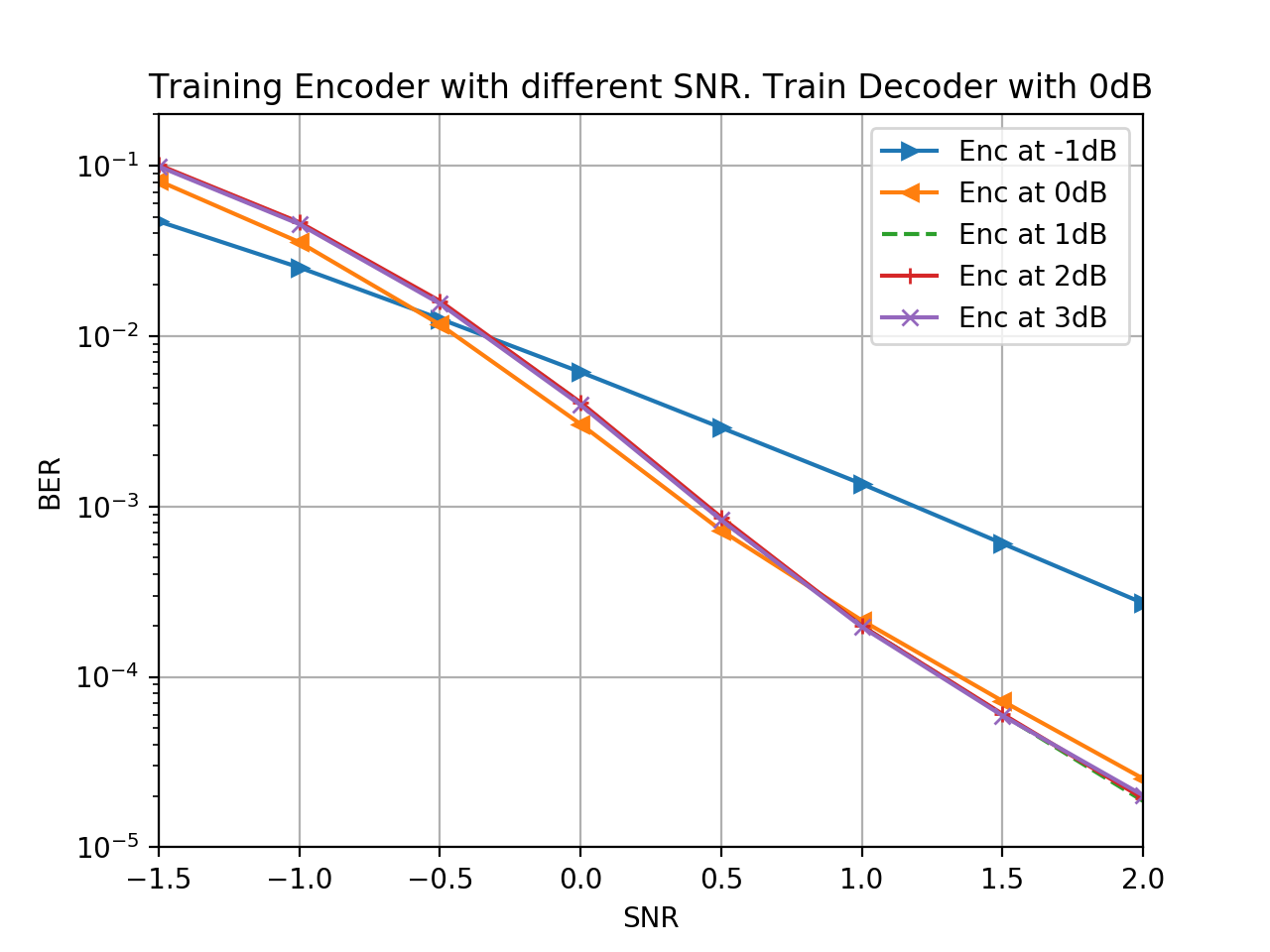}\ \ \ 
\includegraphics[width=0.50\textwidth]{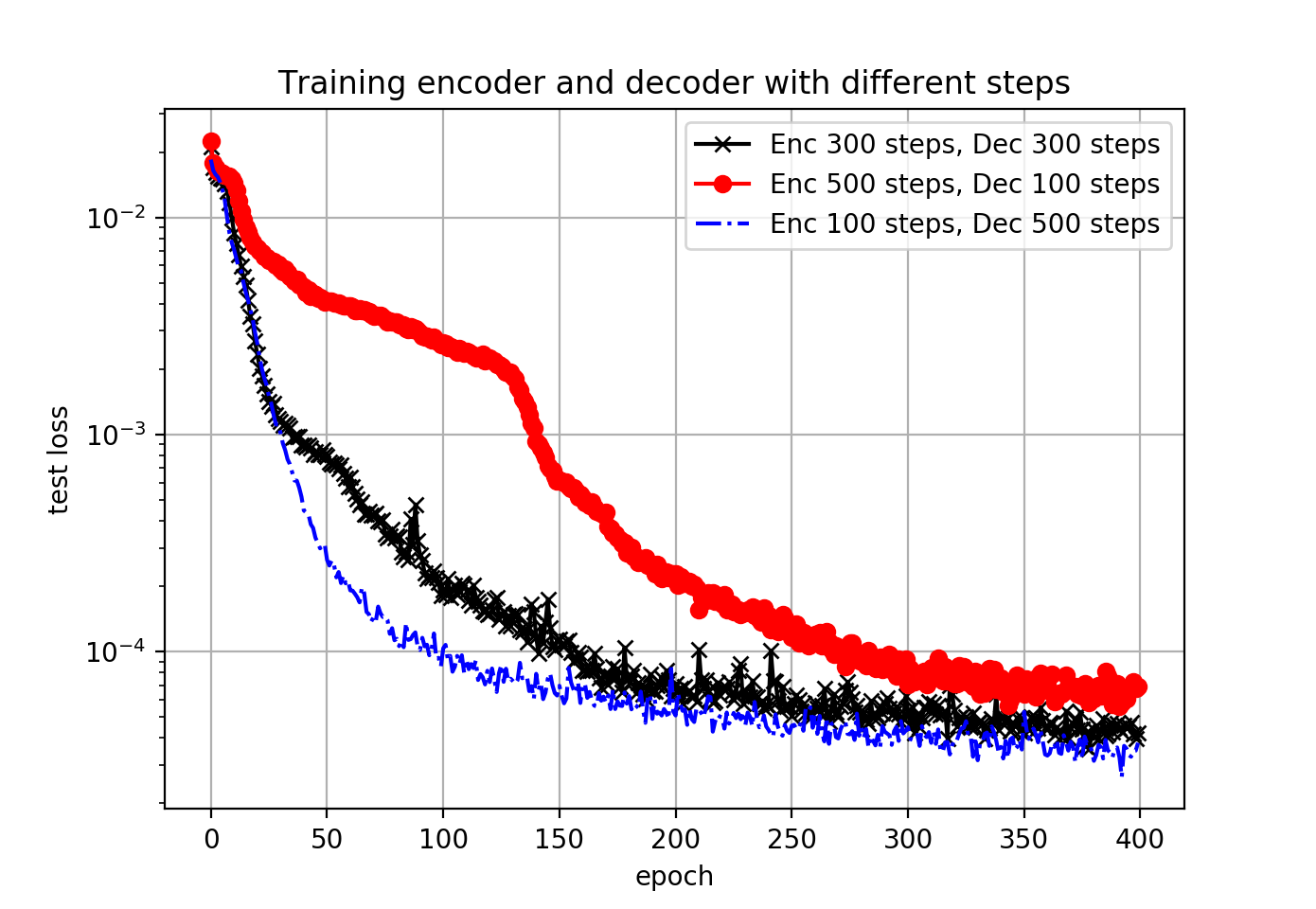}\ \ \ 

\caption{Encoder Training SNR has different coding gain effects (left); Training decoder more lead to faster convergence (right).}\label{train_snr}
\centering
\end{figure}

As the effect of decoder training noise has been studied in~\cite{kim2018communication}, in this section, we study the training SNR of encoder, with fixing decoder training SNR to be 0dB as shown in Figure \ref{train_snr} left. We see that the most reliable code can be learned when training SNR matches testing SNR. Throughout the paper, we make encoder training SNR equals the testing SNR, e.g we testing TurboAE performance at 2dB, we train TurboAE with encoder SNR at 2dB and decoder at 0dB. The BER curve shown in main context is the lower envelope of all curves. 

When encoder training SNR is larger than 1dB (e.g., 1dB, 2dB and 3dB), the BER curves remain nearly the same. Thus encoder training noise level for high SNR region shows diminishing effects on high SNR, which creates an error floor for TurboAE. In main context we state that neural code are suboptimal in high SNR region, since the error is hard to encounter (with probability less than $10^{-4}$), which makes it hard to gather negative examples to train encoder. Improving high SNR region coding gain with data imbalance is an interesting future research direction.

\subsubsection{Train decoder more than encoder}
We argue that when the decoder is well-paired to the fixed encoder, the gradient passed to encoder is more accurate. Training decoder more times will improve performance, as shown in Figure \ref{train_snr} right.  Training decoder more times lead to faster convergence.

\subsubsection{Learning Rate and Batch Size Scheduling}
Increasing batch size improve generalization rather than reducing learning rate~\cite{smith2017don}. To reduce computational expense, we start with batch size $B=500$, and double the batch size when the test loss saturates for 20 epochs till $B=2000$ which is our GPU memory limit. Figure \ref{cnn_vs_rnn} shows that there exists long `fake saturating' points where the test loss saturates for over 20 epochs and then continue to drop. When $B=2000$, when saturates for longer than 20 epochs, the learning rate $lr$ is reduced by 10 times till learning rate reaches $lr=0.000001$.

\subsubsection{Block Error Rate Performance comparison}
\begin{figure}[!h] 
\centering
\includegraphics[width=0.90\textwidth]{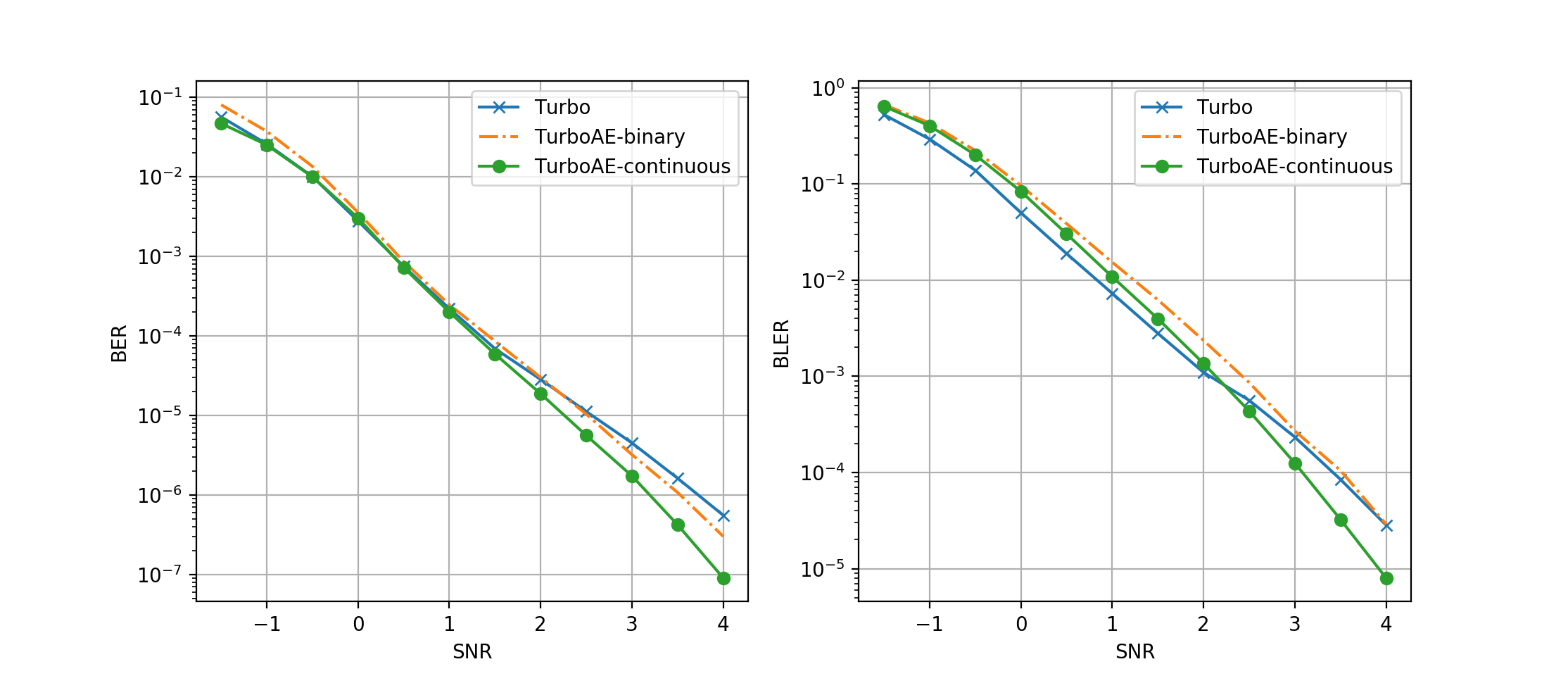}\ \ \ 
\caption{TurboAE BER (left) and BLER (right) performance}\label{ber_bler}
\centering
\end{figure}

The loss function used is Binary Cross-Entropy (BCE), which minimizes average cross entropy for all bits along the block, aiming at minimizing BER. Optimizing BER doesn't necessarily result in optimizing block error rate (BLER), as shown in Figure \ref{ber_bler}. TurboAE-binary shows better performance comparing to Turbo code in BER sense under all SNR points, the BLER performance is worse than Turbo code. 




\section{Complexity Comparison}\label{complex}
Neural networks are known to have high implementation complexity than canonical algorithms. CNN structure is more favorable than RNN since it is of less complexity and easier to run in parallel.

We compare the inference complexity between TurboAE with CNN and GRU implementations (with similar performance), as well as canonical Turbo decoder in this section. The neural network computation is measured via float-point operations (FLOP) in one block. Turbo's encoder and decoder complexity is computed in elementary math operations (EMO), which are are in Table \ref{table:1} : 

\begin{table}[h!]
\tiny{
\centering
\begin{tabular}{|c c c c c c c|} 
 \hline
 Metric & CNN encoder & CNN decoder & GRU encoder & GRU decoder &Turbo encoder & Turbo decoder \\ [0.5ex] 
 \hline
 FLOP/EMO & 1.8M & 294.15M & 334.4M  & 6.7G   &104k &408k\\
 Parameters &152.4k &2.45M & 1.14M & 2.714M & N/A & N/A \\[1ex] 
 \hline
\end{tabular}
\caption{FLOP and number of parameter comparison on block length 100 and 6 iterations}
\label{table:1}
}
\end{table}

CNN encoder and decoder are considered as small, comparing to typical deep learning models which take about 1G FLOP per instance. GRU has much larger FLOP comparing to CNN. Empirically using GRU takes 10x GPU memory and is 10x slower to train. 
However compared to canonical Turbo encoder and decoder, FLOP of TurboAE with CNN is still much larger than canonical decoders. 

We are expecting continuing research would lead to smaller FLOP, as well as the advance of AI-chips will increase the performance when applying CNN to TurboAE.

Due to TurboAE complexity and flexibility, and superior performance on moderate block length on low-to-moderate SNR, the best application area for TurboAE is on dynamical environment (e.g operating on moderate block length and channel with uncertainty) such as low latency code and control plane. On high throughput data plane where canonical codes such as LDPC and Turbo, or neural decoder can be the best method with low complexity and high reliability. In the future, combining both adaptive neural code and human-designed capacity-approaching codes will give more seamless and high reliable communication experience.



\section{TurboAE Performance}
\subsection{Benchmarks}
We use MATLAB-based Vienna 5G simulator and Python-based Commpy~\cite{commpy} as our benchmarks. 

\subsubsection{Vienna 5G simulator}
The detailed implementation details of Vienna 5G are:

\begin{itemize}
\item LDPC code with PWL-Min-Sum decoding algorithm, with 32 decoding iterations. 
\item Polar code with CRC-List-SC decoding algorithm, with list size 32.
\item Turbo code with Linear-Log-MAP decoding algorithm, with 6 decoding iterations. 
\item TBCC code with  MAX-Log-MAP decoding algorithm.
\end{itemize}

TurboAE and Turbo code uses the same number of iterations. Turbo codes simulation results are different between Commpy and Vienna 5G simulator, since Commpy implements vanilla Turbo code, and Vienna 5G simulator implements more advanced coding Turbo schemes. We use Commpy Turbo code as our benchmark. Note that Commpy shows better performance than Vienna 5G, but shows less coding gain on high SNR. We use Commpy as the benchmark, which is the same as~\cite{kim2018communication}.

For Vienna 5G simulator, we find that the code rate for each channel coding is not enforced, e.g when setting code rate $R = 1/3$ with block length $K=100$, the encoder not necessarily outputs codeword with block length $N=300$, but rather outputs longer block length $N=384$. To make a fair comparison, we tune the code rate to enforce the output of encoder to have block length $N=300$, which results in a different setup code rate: 

\begin{itemize}
\item Polar code: for code rate $R=1/2$, the setup code rate is $R= 0.62$; for code rate  $R=1/3$, the setup code rate is $R=0.415$.
\item TBCC Code: for code rate $R=1/2$, the setup code rate $R= 0.64$, for code rate $R=1/3$, the setup code rate $R=0.4275$.
\item Turbo Code: for code rate $R=1/2$, the setup code rate $R=0.62$, for code rate $R=1/3$, the setup code rate $R=0.4175$.
\item LDPC code: for code rate $R=1/2$, the setup code rate $R=0.705$, for code rate $R=1/3$, the setup code rate $R=0.522$.
\end{itemize}

Interested reader can contact Vienna 5G simulator's authors to get access to the code. 

\subsubsection{Commpy on Turbo Code}
RSC code with generating function $(1, \frac{f_1(x)}{f_2(x)})$ is the component code for Turbo code. The generating function of Turbo's RSC affects the performance. Two commonly used configurations of RSC are implemented in Commpy:
\begin{itemize}
    \item code rate $R=1/3$, with $f_1(x) = 1 + x^2$ and $f_2(x) = 1+x+x^2$, which is denoted as Turbo-757.
    \item code rate $R=1/3$, with $f_1(x) = 1 + x^2 +x^3$ and $f_2(x) = 1+x+x^3$, which is standard Turbo code used in LTE system, denoted as Turbo-LTE.
\end{itemize}

\begin{figure}[!h] 
\centering
\includegraphics[width=0.95\textwidth]{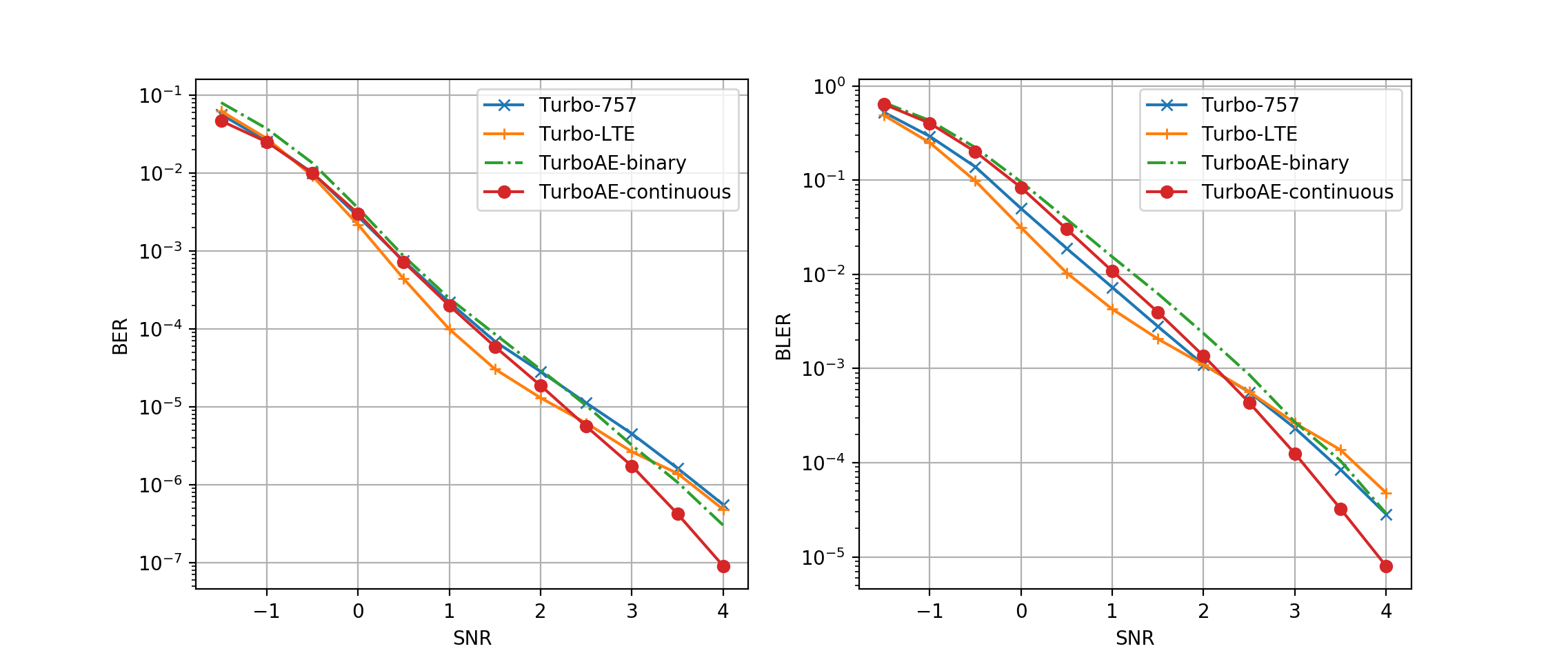}\ \ \ 
\caption{Commpy TurboAE with different trellis performance BER (left), and BLER (right)}\label{commpy}
\centering
\end{figure}

In main context, the benchmarks are using with Turbo-757, while the performance comparison between Turbo-757 and Turbo-LTE are shown in Figure~\ref{commpy}. The performance trend are the same, while Turbo-LTE shows slightly better performance. The same claim in main text on Turbo-757, works for Turbo-LTE.

\subsection{Continuous Channels}
Continuous channel refers to the channel where the received signal $y$ can is continuous, where both TurboAE-continuous and Turbo-binary can be supported. We discussed short block performance on AWGN and non-AWGN channels, and in this section we discuss the longer block length, and other channels.

\subsubsection{Scale to Long Code Block Length is hard}

Figure~\ref{fading_bl100} left shows that after fine-tuning at block length 1000, fine-tuned TurboAE shows improved performance comparing to TurboAE trained on block length 100 and tested on block length 1000. However, TurboAE-continuous shows worse performance comparing to canonical Turbo code. As shown in main context, TurboAE's coding gain on long block length is smaller than Turbo code due to trainability and computation issues. Improving performance on long block length is an interesting future research direction.

\subsubsection{TurboAE on iid Rayleigh Fading Channel}
\begin{figure}[!h] 
\centering
\includegraphics[width=0.95\textwidth]{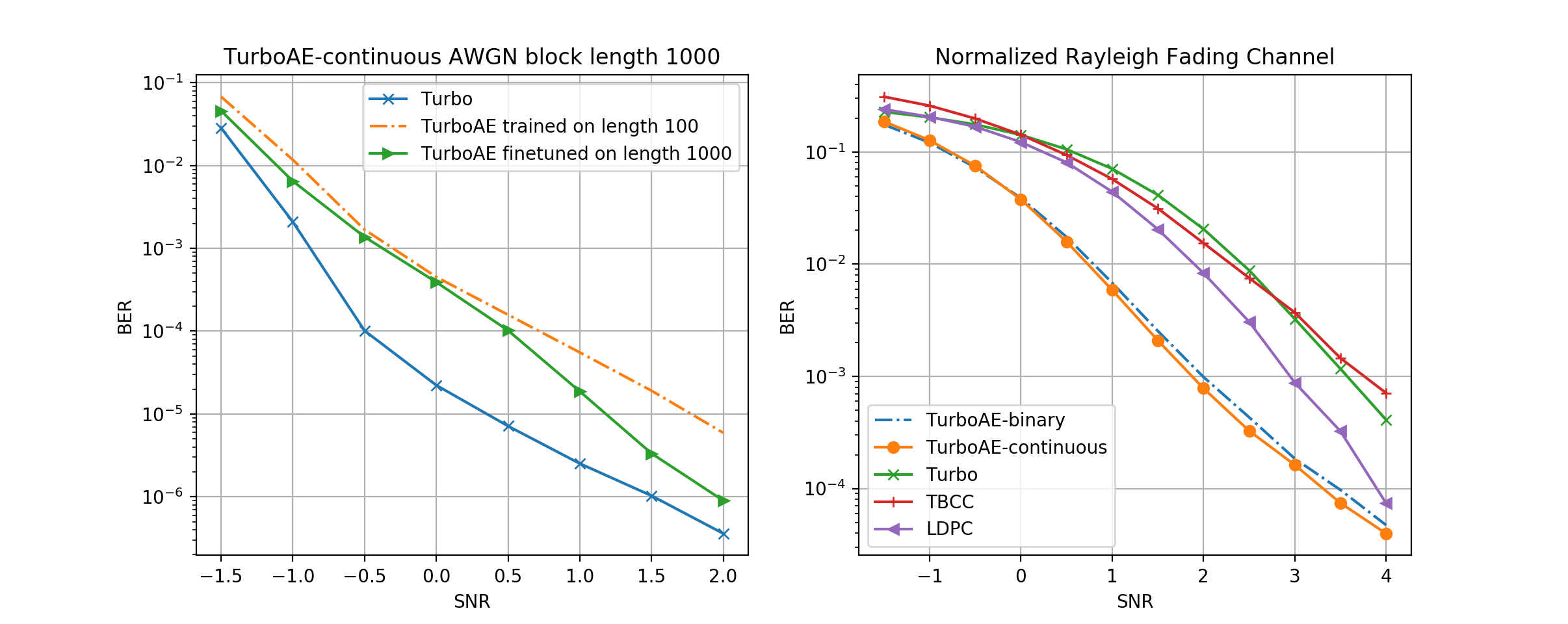}\ \ \ 
\caption{TurboAE performance on block length 1000 (left) and TurboAE on Rayleigh Fading Channel (right)}\label{fading_bl100}
\centering
\end{figure}

Non-coherent Rayleigh Fading Channel is defined as $y_i = h_i x_i + z_i$, where iid $z_i \sim N(0, \sigma^2)$, and $h_i$ is normalized iid Rayleigh distribution fading noise as $h_i \sim \frac{\sqrt{U^2+V^2}}{\sqrt{\pi/2}}$, while $U$ and $V$ are IID unit Gaussian variables. Non-coherent setting means the decoder doesn't know $h_i$: the benchmarks (canonical decoders include Turbo, TBCC, and LDPC) are not aware of the fading component by still taking log-likelihood as decoder input, while TurboAE is not further trained to learn $h_i$. The performance of Non-coherent Rayleigh Fading Channel are shown in Figure \ref{fading_bl100} right. On Non-coherent Rayleigh Fading Channel, TurboAE-binary and TurboAE-continuous outperforms LDPC, TBCC and Turbo code in a wide SNR region. 

\subsubsection{TurboAE-continuous combines Modulation and Coding}

In the main context, we show that on ATN channel, TurboAE-continuous outperforms TurboAE-binary. 
TurboAE-continuous outperforms TurboAE-binary significantly at high SNR since TurboAE-continuous jointly learns modulating and coding in continuous value domain, which has better advantage at high SNR schemes. 

\begin{figure}[!h] 
\centering
\includegraphics[width=0.5\textwidth]{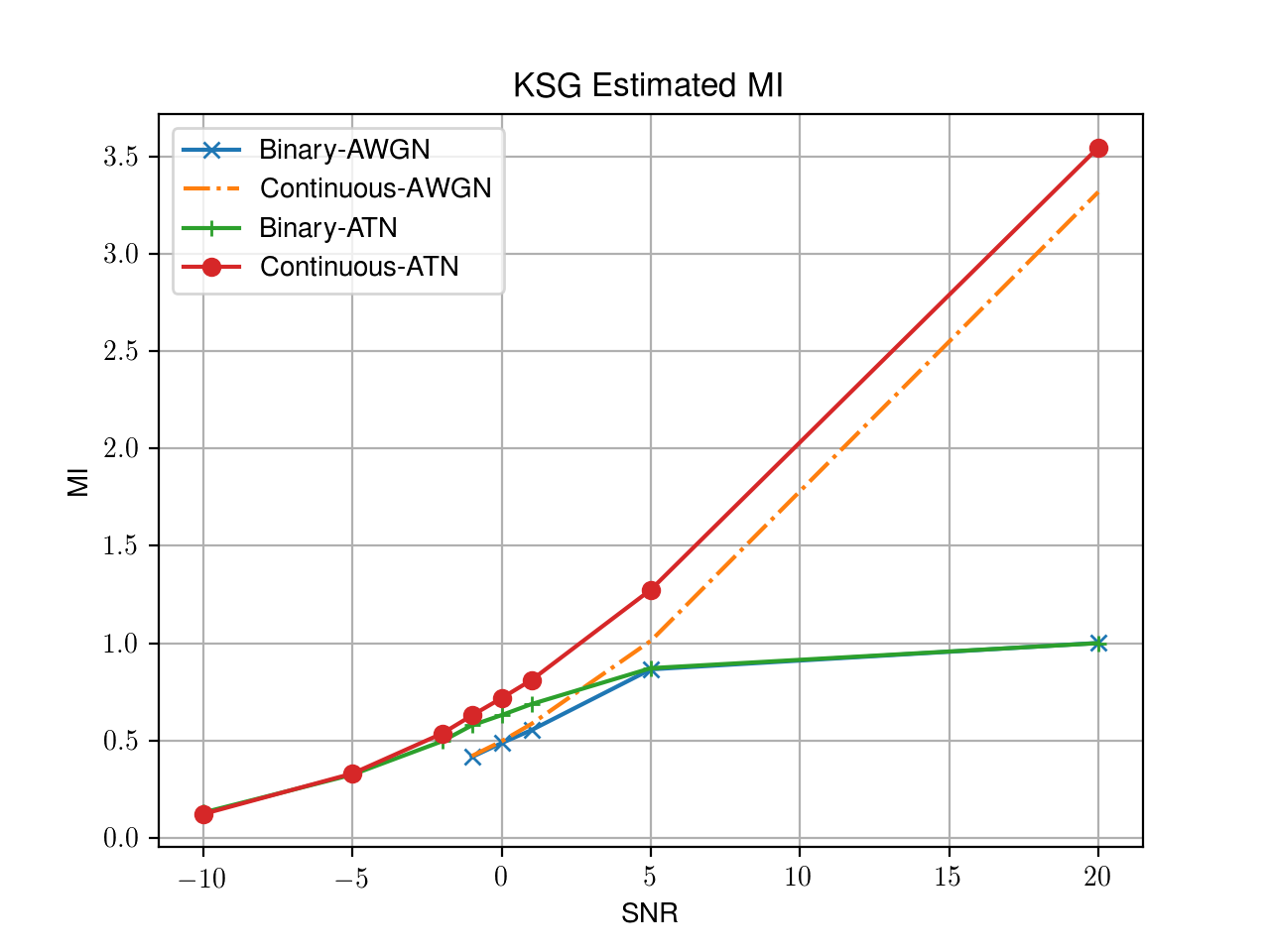}\ \ \ 
\caption{KSG estimated Mutual Information for AWGN and ATN channel}\label{ksg}
\centering
\end{figure}

To investigate the fundamental coding gain of TurboAE-continuous in high SNR schemes, we investigate the channel capacity of non-AWGN channel (take ATN as example) and AWGN channel, as shown in Figure~\ref{ksg} right. Binary-AWGN and Continuous-AWGN refers to the channel capacity where code $x$ is binary and continuous at AWGN channel, respectively. Binary-ATN and Continuous-ATN refers to the channel capacity where code $x$ is binary and continuous at ATN channel. We use estimated Mutual Information (MI) via KSG estimator~\cite{kraskov2004estimating} as the surrogate measure for channel capacity, as there is no close-form channel capacity for ATN channel. 

Under the SNR range where most channel coding operates around (0dB), the MI between Binary-AWGN and Continuous-AWGN is very close, thus applying continuous coding doesn't improve coding gain significantly on AWGN channel. However, the MI between binary-ATN and Continuous-ATN is significant, thus applying continuous code can further increase the channel capacity comparing to using binary code on non-AWGN channel. Moreover, at high SNR, the capacity of continuous code is much larger than binary code, which shows that Turbo-AE, as a method to learn continuous code, has theoretical advantage on high SNR schemes. 

\subsection{Binary Channels}

Binary channels restrict the decoder input to be binary, which only supports binary operations. Only TurboAE-binary is supported. We use the following canonical binary channels:

\begin{itemize}
\item iid Binary Symmetric Channel (BSC), $x \in \{-1, +1\}$ and $y \in \{-1, +1\}$,  flip rate $P(y \neq x) = p_{bsc}$, and $P(y = x) = 1 - p_{bsc}$.
\item iid Binary Erasure Channel(BEC), $x \in \{-1, +1\}$ and $y \in \{-1, 0, +1\}$, while $y=0$ represents erasure. Erasure rate $P(y = 0) = p_{bec}$, and $P(y = x) =1 -  p_{bec}$.
\end{itemize}

\begin{figure}[!h] 
\centering
\includegraphics[width=0.95\textwidth]{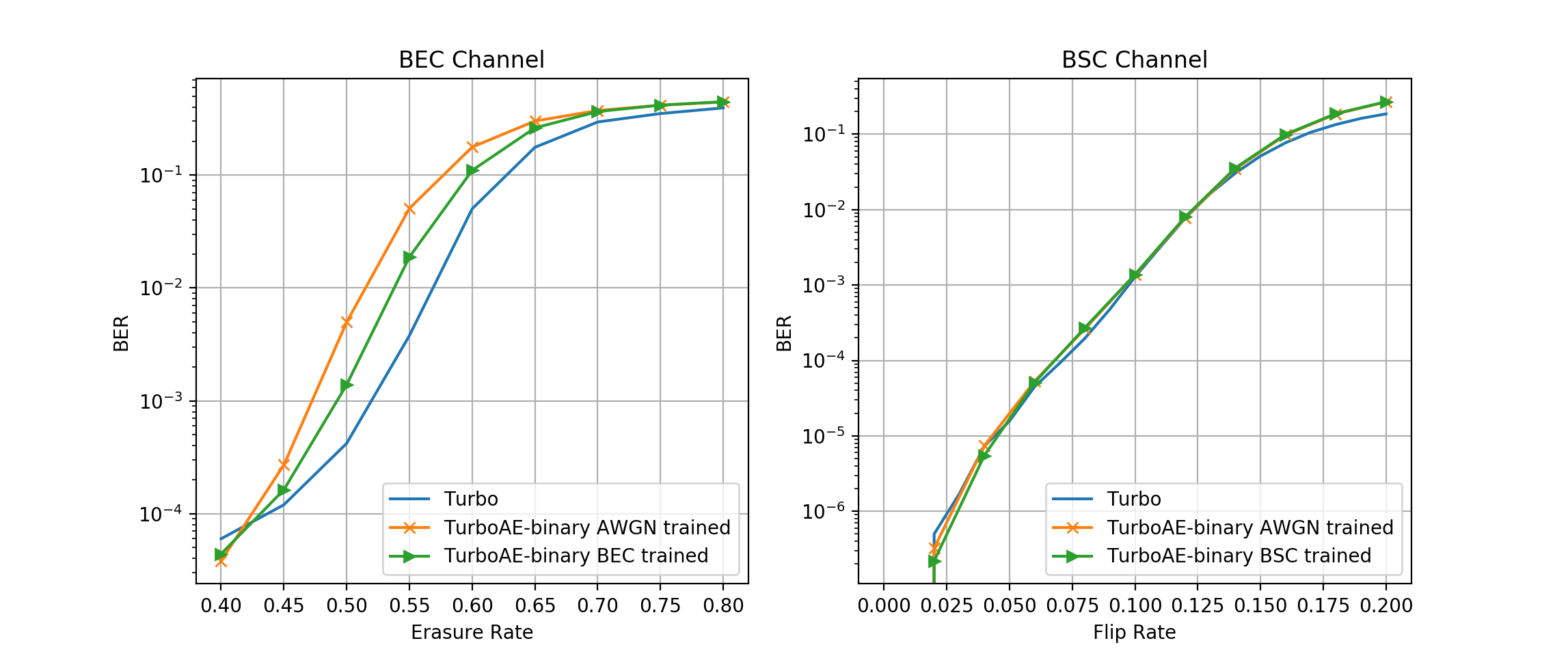}\ \ \ 
\caption{BEC and BSC performance}\label{bsc_bec}
\centering
\end{figure}
On BSC channel, TurboAE-binary and Turbo works nearly the same, which implies that AWGN-trained TurboAE can generalize to BSC channel. 

On BEC channel, TurboAE-binary trained on AWGN works worse than Turbo, since on AWGN channel there doesn't exist erasure. However TurboAE fine-tuning on BEC channel still has gap comparing to Turbo. The result shows that the trainability of TurboAE still needs improvement.

\subsection{Interleaved Encoding Visualization}
\begin{figure}[!h] 
\centering
\includegraphics[width=0.5\textwidth]{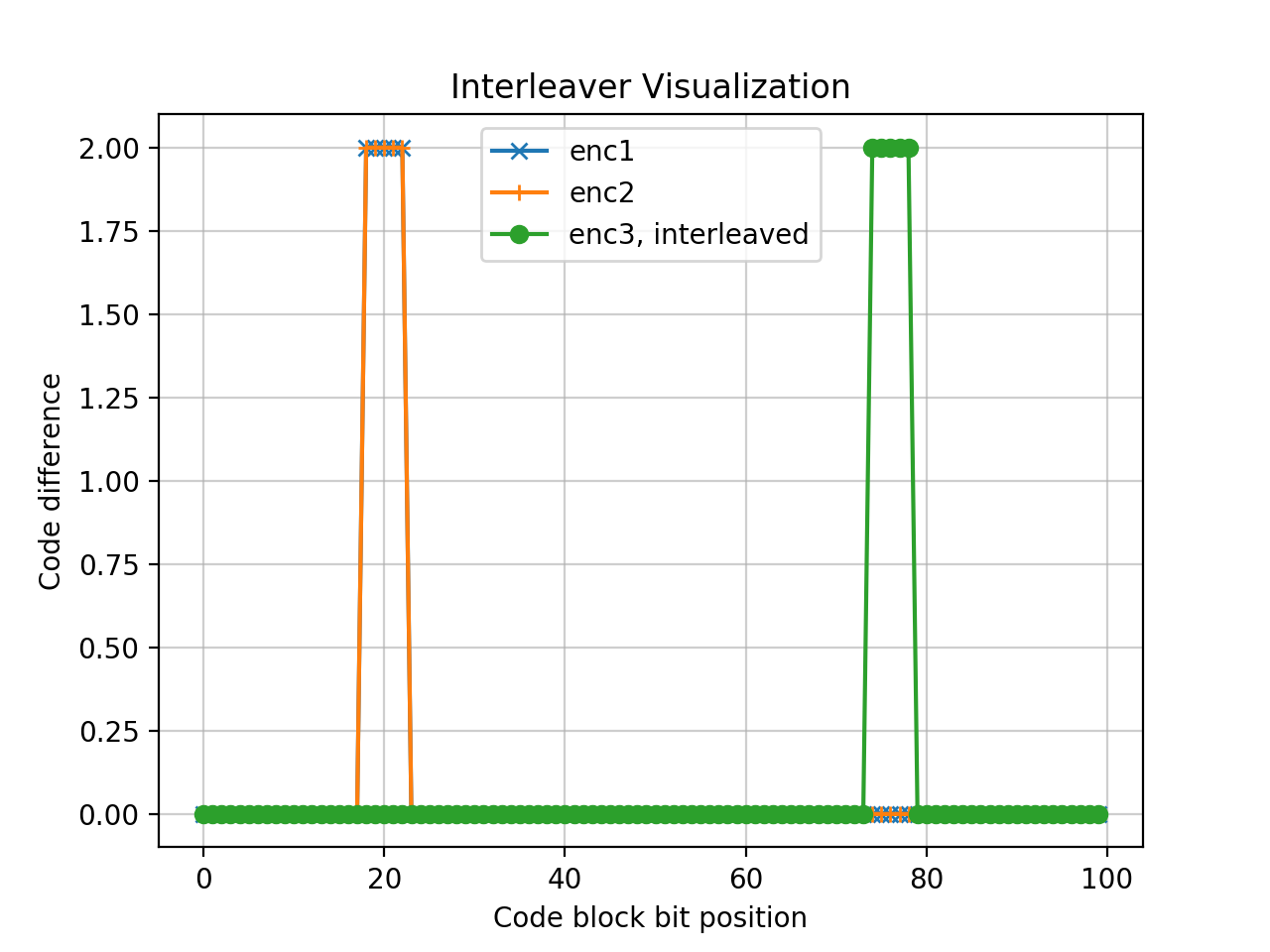}\ \ \ 
\caption{Randomness added via interleaving}\label{vis_int}
\centering
\end{figure}

We test the random coding effect of interleaved encoder with 2 same message, $u_1$ and $u_2$, and perturb at position index 20, which makes the only difference between $u_1$ and $u_2$ is $s_1[20] = 1.0$ and $s_1[20] = 0.0$. We plot the absolute maximized code difference $|f_\theta(s_1) - f_\theta(s_2)|$ for all three encoding blocks. With interleaved encoder, one single message bit change (at code bit 20) can cause random non-adjacent bits (at code bit 75) to change, which adds encoding randomness, shown in Figure~\ref{vis_int}.

\end{document}